\documentclass[preprint2]{aastex}
\usepackage{natbib}

\def\lesssim{\mathrel{\hbox{\rlap{\hbox{%
 \lower4pt\hbox{$\sim$}}}\hbox{$<$}}}}

\begin{document}

\title{Physical Properties of the X-ray Luminous SN~1978K in NGC 1313 from
Multiwavelength Observations\footnote{Based upon (i) archival {\it ASCA}
observations obtained from the High Energy Astrophysics Science
Archive Research Center On-line Service, which is provided by
NASA-Goddard Space Flight Center, (ii) {\it ROSAT} HRI PI
observations, (iii) observations with the NASA/ESA {\it Hubble Space
Telescope} (Faint Object Spectrograph), obtained at the Space
Telescope Science Institute, which is operated by Associated
Universities for Research in Astronomy, Inc., under NASA contract
NAS5-26555, and (iv) observations using the Australia Telescope
Compact Array, which is funded by the Commonwealth of Australia for
operation as a National Facility managed by CSIRO.}}

\author{Eric M. Schlegel\altaffilmark{2}, Stuart
Ryder\altaffilmark{3}, L. Staveley-Smith\altaffilmark{4},
R. Petre\altaffilmark{5}, E. Colbert\altaffilmark{5,6},
M. Dopita\altaffilmark{7}, D. Campbell-Wilson\altaffilmark{8}}

\altaffiltext{2}{High Energy Astrophysics Division, Smithsonian
Astrophysical Observatory, Harvard-Smithsonian Center for
Astrophysics, Cambridge, MA 02138}

\altaffiltext{3}{Joint Astronomy Center, 660 N. A'Ohoku Place, Hilo, HI 96720}

\altaffiltext{4}{Australia Telescope National Facility, CSIRO, P.O. Box 76, Epping, NSW 2121, Australia}

\altaffiltext{5}{X-ray Astrophysics Group, Laboratory for High Energy Astrophysics, NASA-GSFC, Greenbelt, MD 20771}

\altaffiltext{6}{National Research Council Fellow}

\altaffiltext{7}{Mount Stromlo and Siding Spring Observatories, Australian National University, Weston Creek P. O., ACT 2611, Australia}

\altaffiltext{8}{Astrophysics Dept., School of Physics A29, University of Sydney, Sydney NSW 2006, Australia}

\begin{abstract}

We update the light curves from the X-ray, optical, and radio
bandpasses which we have assembled over the past decade, and present
two observations in the ultraviolet using the {\it Hubble Space
Telescope} Faint Object Spectrograph.  The HRI X-ray light curve is
constant within the errors over the entire observation period.  This
behavior is confirmed in the {\it ASCA} GIS data obtained in 1993 and
1995.  In the ultraviolet, we detected Ly-${\alpha}$, the [Ne~IV]
2422/2424 {\AA} doublet, the Mg II doublet at 2800 {\AA}, and a line
at $\sim$3190 {\AA} we attribute to He I 3187.  Only the Mg II and He
I lines are detected at SN1978K's position.  The optical light curve
is formally constant within the errors, although a slight upward trend
may be present.  The radio light curve continues its steep decline.

The longer time span of our radio observations compared to previous
studies shows that SN1978K is in the same class of highly X-ray and
radio-luminous supernovae as SN1986J and SN1988Z.  The [Ne~IV]
emission is spatially distant from the location of SN1978K and
originates in the pre-shocked matter.  The Mg II doublet flux ratio
implies the quantity of line optical depth times density of
$\sim$10$^{14}$ cm$^{-3}$ for its emission region.  The emission site
must lie in the shocked gas.

\end{abstract}

\keywords{stars: supernovae: individual: SN1978K}

\section{Introduction}

SN~1978K was the second supernova detected and recognized as a
supernova from its radio emission and the first from its X-ray flux
\citep[Paper I]{R+93}.  It was discovered in a {\it ROSAT} PSPC
observation of NGC~1313 but had not been detected by an {\it Einstein}
observation of the same field 11 years earlier.

SN~1978K is one of a handful of extremely luminous, X-ray emitting
supernovae \citep{S95}.  X-ray observations are difficult to obtain,
so to date, few observations exist.  Currently, nine supernovae are
known to emit, or have emitted, X-rays sometime in the months and
years following their outbursts (SN~1978K, SN~1979C, SN~1980K,
SN~1986J, SN~1987A, SN~1988Z, SN~1993J, SN~1994I, SN~1995N (references
are listed below).  Of these, the X-rays from SN~1987A largely
resulted from Compton scattered ${\gamma}$-rays from the radioactive
decay of $^{56}$Co.  SN~1987A is very likely starting to undergo a
circumstellar interaction \citep{HAT96}; it is expected to become a
very bright X-ray source \citep{CD95,MN94,SSN93}.

\begin{figure*}
\caption[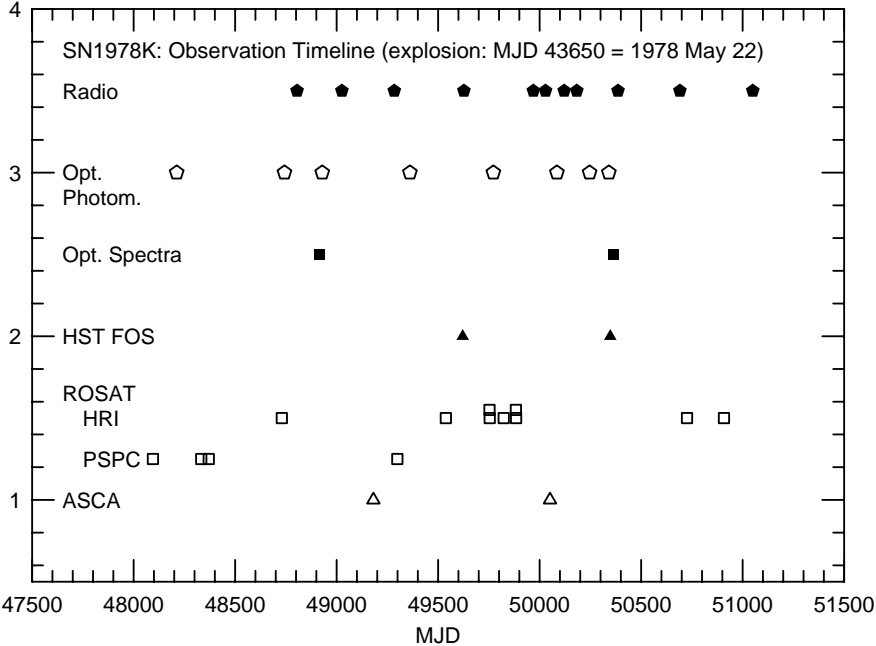]{The distribution of all observations described
in this paper.  The vertical axis is an arbitrary index used to spread
out the observations but the observations are organized by energy from
high (bottom) to low (top).  The first of the two optical spectra was
obtained by and described in \cite{CDD95}.  See Tables 1, 2, 3, 6,
and 8 for the details of each observation.} \label{obs_plot}
\rotatebox{-90}{\scalebox{0.5}{\includegraphics{sn78k_fig1.ps}}}
\end{figure*}

The remaining eight supernovae are all believed to emit X-rays solely
because of a circumstellar interaction. SN~1980K \citep{CKF82} and
SN~1993J \citep{Zim94} were detected in the days immediately following
their explosions.  SN~1979C \citep{IPA98}, SN~1988Z \citep{FT96},
SN~1994I \citep{IPA298}, and SN~1995N \citep{LZB96} have all been
recently detected so that we do not as yet have any X-ray light
curves.  Just three supernovae remain: SN~1978K, SN~1986J, and
SN~1993J.  SN~1986J and SN~1993J have been reported to be fading
\citep{HBCT98,Zim94}; the X-ray light curve of SN~1978K appeared to be
constant \citep[SPC96]{SPC96} as of 1995.

Modeling of the radio spectrum by \cite[MWP]{MWP} led them to infer
the existence of absorption by an H\,{\sc ii}~region along the line of
sight to SN~1978K. Based on an optical spectrum of SN~1978K taken in
1992~October, \cite{CDD95} hypothesised that the optical and X-ray
emission is driven by shock waves propagating through a clumpy
circumnuclear wind, rather than through the extremely massive
circumstellar envelope invoked in Paper~I. In a recent paper,
\cite{Chu99} present a spectrum of the H$\alpha$ + [N\,{\sc ii}]
region at high resolution, which shows broad emission from the
supernova ejecta, with superimposed (narrow) nebular lines. They
attribute the latter to a circumstellar ejecta nebula from the
progenitor of SN~1978K, which the supernova ejecta could plough into
quite soon. This would cause significant brightening of SN~1978K at
all wavelengths, making continued monitoring all the more important.

In this paper, we update the X-ray light curve with additional {\it
ROSAT} HRI observations that have been obtained since SPC96, briefly
discuss the two epochs of {\it ASCA} data, update the optical and
radio light curves, and present {\it HST} FOS spectra.  Throughout
this paper, we assume the distance to NGC 1313 is 4.5 Mpc
\citep{deV63}; the corresponding image scale is 1'' = 22 pc.

\section{Observations}

For a multi-wavelength paper such as this one, we present the details
of the observations and the analyses (next section) of the data for
each bandpass in separate subsections in case the reader chooses to
skip some of the details.  Each section is explicitly labeled for
easier navigation.  Figure~\ref{obs_plot} shows the distribution of
all of the observations to be described.

\subsection{X-ray}

\begin{figure*}
\caption[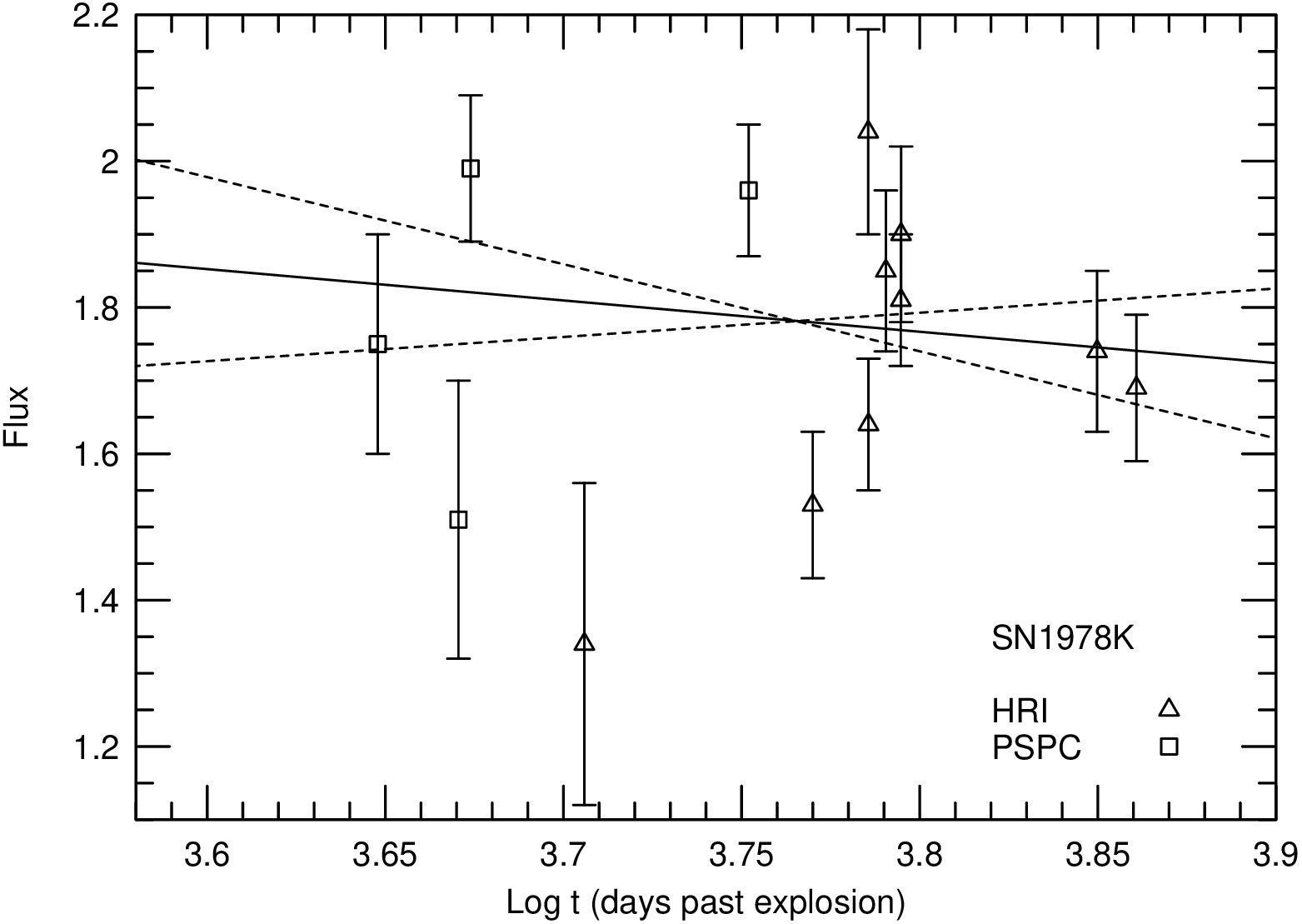]{The {\it ROSAT} X-ray light curve in the 0.2-2.0
keV band for SN~1978K.  The dashed lines show the $\pm$90\% range of
the slope.  The vertical axis is in units of 10$^{-12}$ ergs s$^{-1}$
cm$^{-2}$. See Table~\ref{Obs_table} for details.} \label{HRI_LC}
\scalebox{0.4}{\includegraphics{sn78k_fig2.ps}}
\end{figure*}

\subsubsection{ROSAT Observations}

An early portion of the X-ray light curve of SN~1978K was presented in
\cite[SPC96]{SPC96}.  Since that paper, we have accumulated another
six observations of the supernova which doubles the number of data
points and increases the coverage to 7 years, representing about one
third of the life of SN~1978K. Table~\ref{Obs_table} summarizes the
observations where we also include the PSPC observations (SPC96). The
light curve therefore shows all of the data of SN1978K obtained by
{\it ROSAT}.  The 1998 April observation is among the last
observations to be obtained with {\it ROSAT}.

Recently, \cite{Sn98} analyzed the particle background of the {\it
ROSAT} HRI.  The issues raised in that paper could affect the X-ray
light curve of SN~1978K if the particle background has not been
properly subtracted.  In the light of that paper, the entire HRI data
set of SN~1978K has been re-processed using the tools described in
\cite{Sn98}.

We extracted the counts for each epoch using an aperture of radius
$40''$ which contains about 92\% of the encircled energy
\citep{Da+97}.  The counts from all of the observations were converted
to a flux by adopting the best-fit model that describes the {\it
ROSAT} PSPC and {\it ASCA} SIS data \citep{Pet94a}.  The model
combines a soft (kT$\sim0.84$~keV) thermal component and either a hard
(kT$\sim4.7$~keV) thermal component or a power law ($\Gamma \sim
1.08$). None of the HRI points show evidence of a spectral change as
measured by a hardness ratio; the HRI's ability to discern spectral
changes is, however, quite limited, so the lack of a spectral change
is not restrictive \citep{Pr96}.  The final light curve
(Figure~\ref{HRI_LC}) does not differ from the light curve presented
in SPC96 in any significant or systematic manner.

\subsubsection{ASCA Spectra}

\begin{figure*}
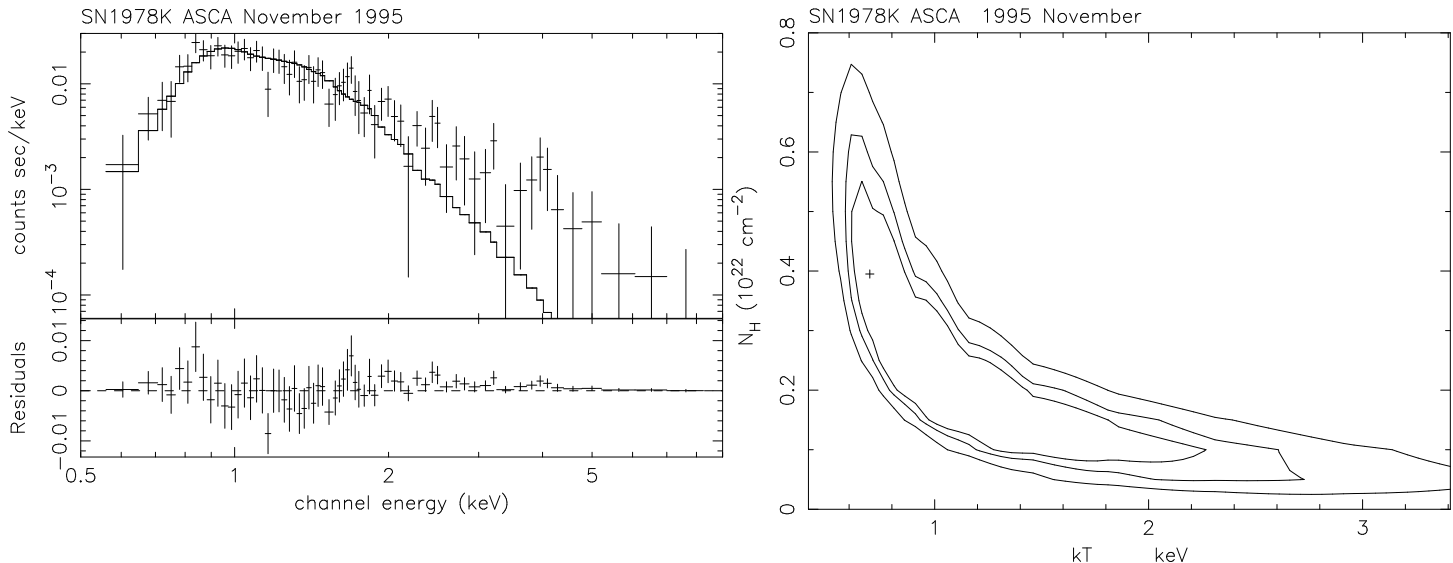

\caption[sn78k_fig3a.ps,sn78k_fig3b.ps]{(a) The extracted {\it ASCA}
spectrum for 1995 November fitted with a hot diffuse gas model.  The
model details are listed in Table~\ref{asca_fits}.  (b) The contours
for the fit to the {\it ASCA} spectrum of 1995 November using a hot
diffuse gas model.  The model details are listed in
Table~\ref{asca_fits}.  The 1993 July contours are similar in shape
and location \cite{Pet94b}.} \label{asca_spec}
\rotatebox{-90}{\scalebox{0.4}{\includegraphics{sn78k_fig3a.ps}}}
\rotatebox{-90}{\scalebox{0.4}{\includegraphics{sn78k_fig3b.ps}}}
\end{figure*}

Two observations of SN1978K have been obtained with the gas
proportional counter (GIS) and the CCD detectors (SIS) on-board {\it
ASCA}, in 1993 July (during the verification phase) and in 1995
November (Table~\ref{asca_table}).  The 1993 observation has been
described in \cite{Pet94b}.  In that paper, the authors did not
detect any spectral lines.  The 1995 observation has not been
published elsewhere, to the best of our knowledge.

We copied the screened event lists for both epochs of data from the
{\it ASCA} archive\footnote{These data were obtained through the High
Energy Astrophysics Science Archive Research Center Online Service
provided by NASA/Goddard Space Flight Center}.  The use of the
screened lists ensures that both data sets have been filtered
identically.  We extracted spectra from each data set taking care to
avoid the other sources in NGC 1313.  The background was obtained from
the same SIS chip or GIS field.  Spectra were extracted for each
detector separately (GIS2, GIS3; SIS0, SIS1) and then added together
to form one GIS and one SIS spectrum per epoch.  Response matrices for
each detector were also summed.  The spectrum for 1995 November is
shown in Figure~\ref{asca_spec}; the model fits will be described in
section~\ref{asca_analy}.

\subsection{HST FOS Observations}

\begin{figure*}
\caption[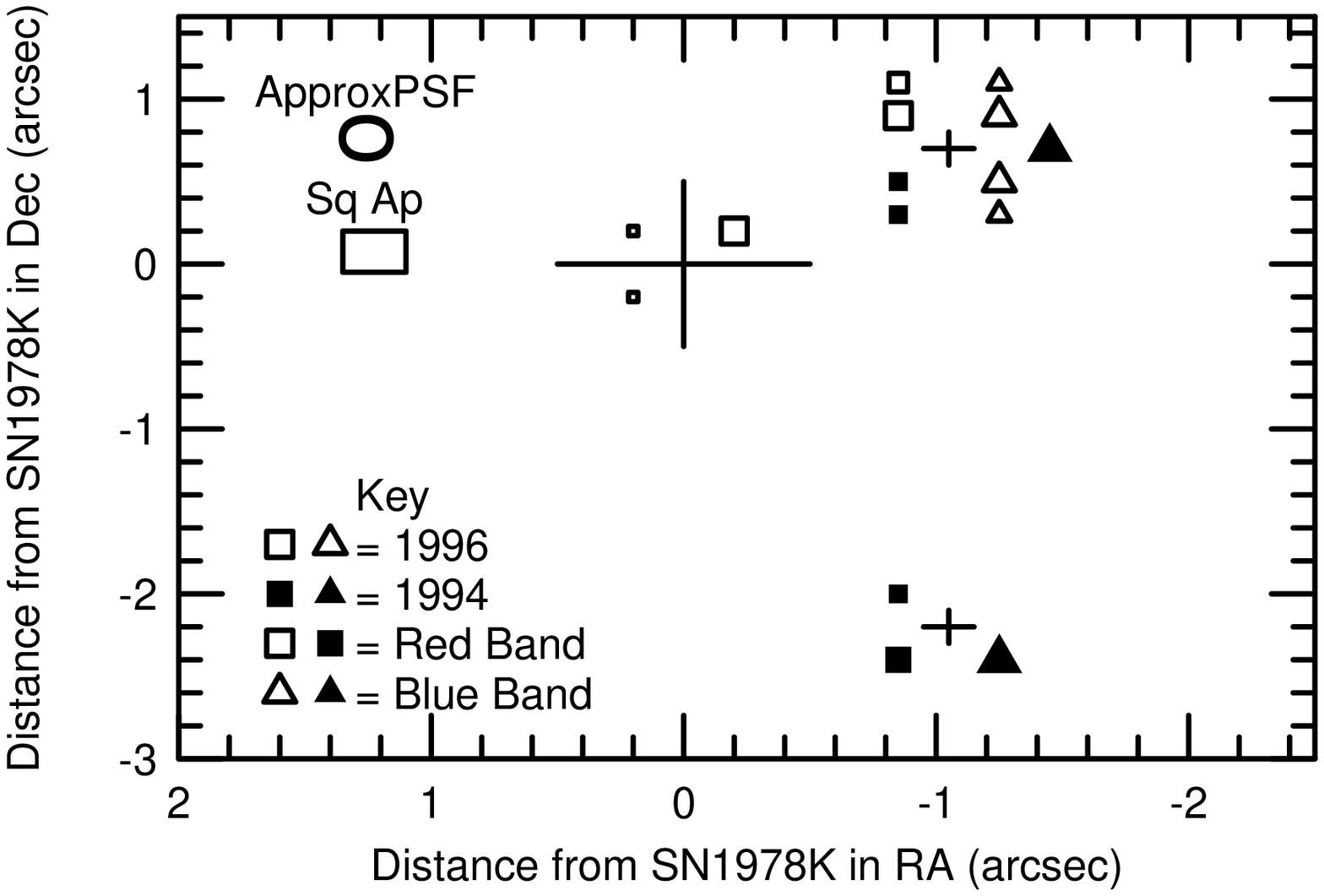]{The complete pointing history (crosses) of the
{\it HST} FOS observations with symbols offset from the crosses
horizontally and vertically for clarity.  The crosses represent the
actual pointing locations; the larger cross is the correct position of
SN1978K.  The symbols are coded as listed in the key.  The size of the
square or triangle crudely represents the exposure time at that
location for the individual observations.  The rectangle in the upper
left corner represents the square FOS aperture; the circular aperture
would be about the same size if it circumscribes the square.  The
approximate 70\% HST encircled energy function is also shown.}
\scalebox{0.6}{\includegraphics{sn78k_fig4.ps}}
\label{point_hist_fig}
\end{figure*}

An observation using the Faint Object Spectograph onboard the {\it
Hubble Space Telescope} was obtained in 1994 September 26 using the
best available radio and optical coordinates.  Target lock failed,
largely because the source is an emission line object.  This accounts
for the large number of off-source pointings. The observation was
rescheduled for 1996 September 22 after improved offset coordinates
were obtained from short WFPC2 exposures (next section).  Some of the
pointing problems stem from the difficulties in tying together the
optical and radio coordinate systems in the southern hemisphere
\citep{R+93}.

The SN1978K observation was designed to obtain on-source ``blue''
($\sim$1200-2500 {\AA} using grating G160L) and ``red''
($\sim$2200-3200 {\AA} using grating G270H) spectra plus an off-source
spectrum for each band.  The approximate boundary between the two
bands lies at $\sim$2350 {\AA} with an overlap of $\sim$100 {\AA}.

\begin{figure*} 
\caption[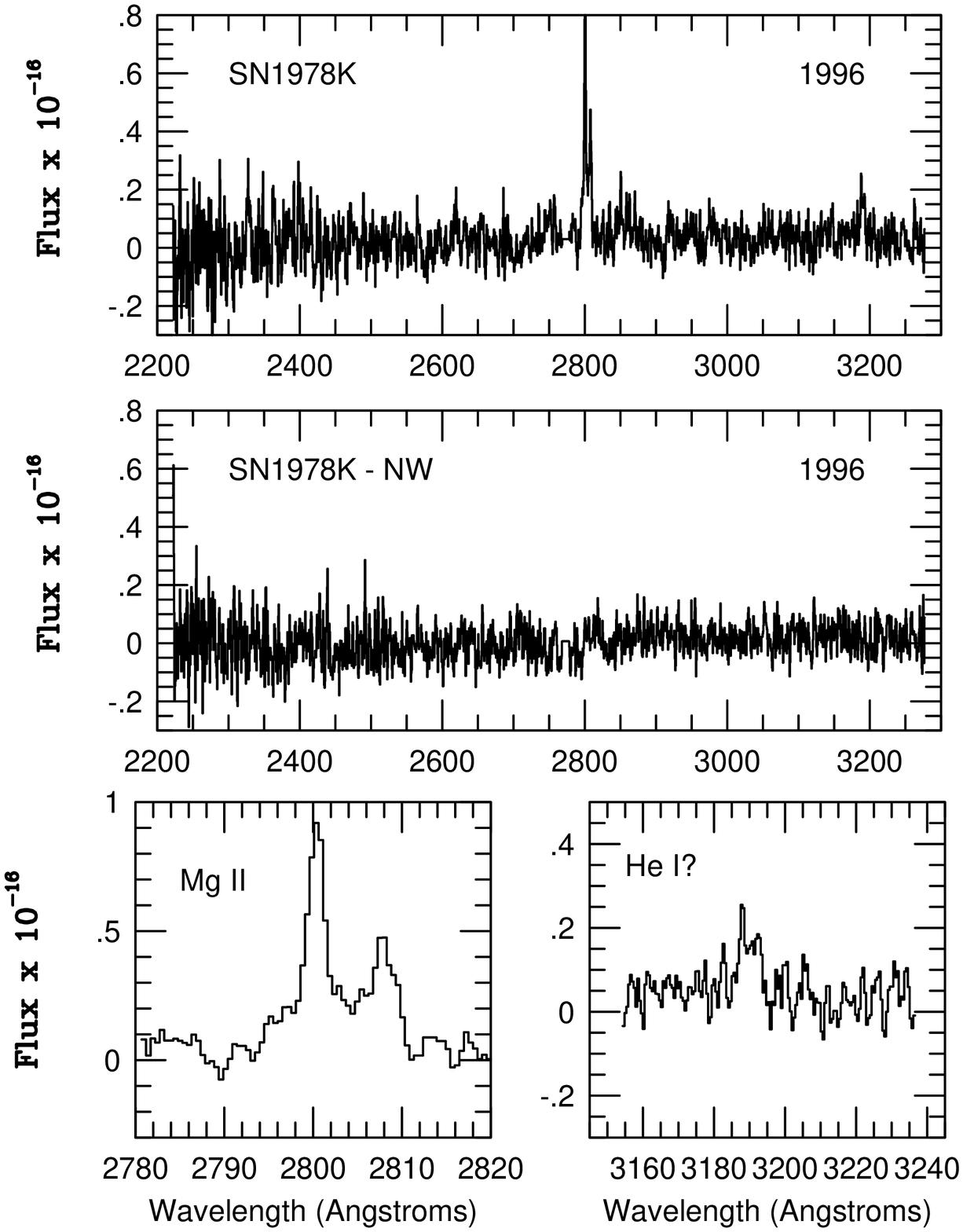]{{\it HST} FOS spectra at SN1978K. The top plot
shows the 1996 red spectrum at the position of SN~1978K; the middle
plot shows the spectrum obtained 1".5 away to the NW.  The spectra
have not been corrected for reddening nor for the redshift of SN~1978K
(439 km s$^{-1}$).  Note that both the top and middle spectra show a
small flat excision, just blueward of the Mg II line, where a noisy
pixel was located.  The bottom spectra expand the regions around the
Mg II line at 2800{\AA} (left) and the line at 3190{\AA} (right).}
\scalebox{0.6}{\includegraphics{sn78k_fig5.ps}}
\label{HST_spec_on}
\end{figure*}

The pointing history, constructed from the ``RA\_APER'' and
``DECAPER'' keywords in the file headers, is shown in
Figure~\ref{point_hist_fig} where the weight of a data point
represents the exposure time at that location.  The total observation
set is described in Table~\ref{point_hist_tab}.  From the pointing
history, we distinguish 7 separate observations: a 1996 red band
observation centered on SN1978K, 1996 blue and red band spectra NW of
SN1978K, 1994 blue and red spectra NW of the SN, and 1994 blue and red
spectra to the SW.  We distinguish between the 1994 and 1996
observations at a given location because the source may vary.

\begin{figure*}
\caption[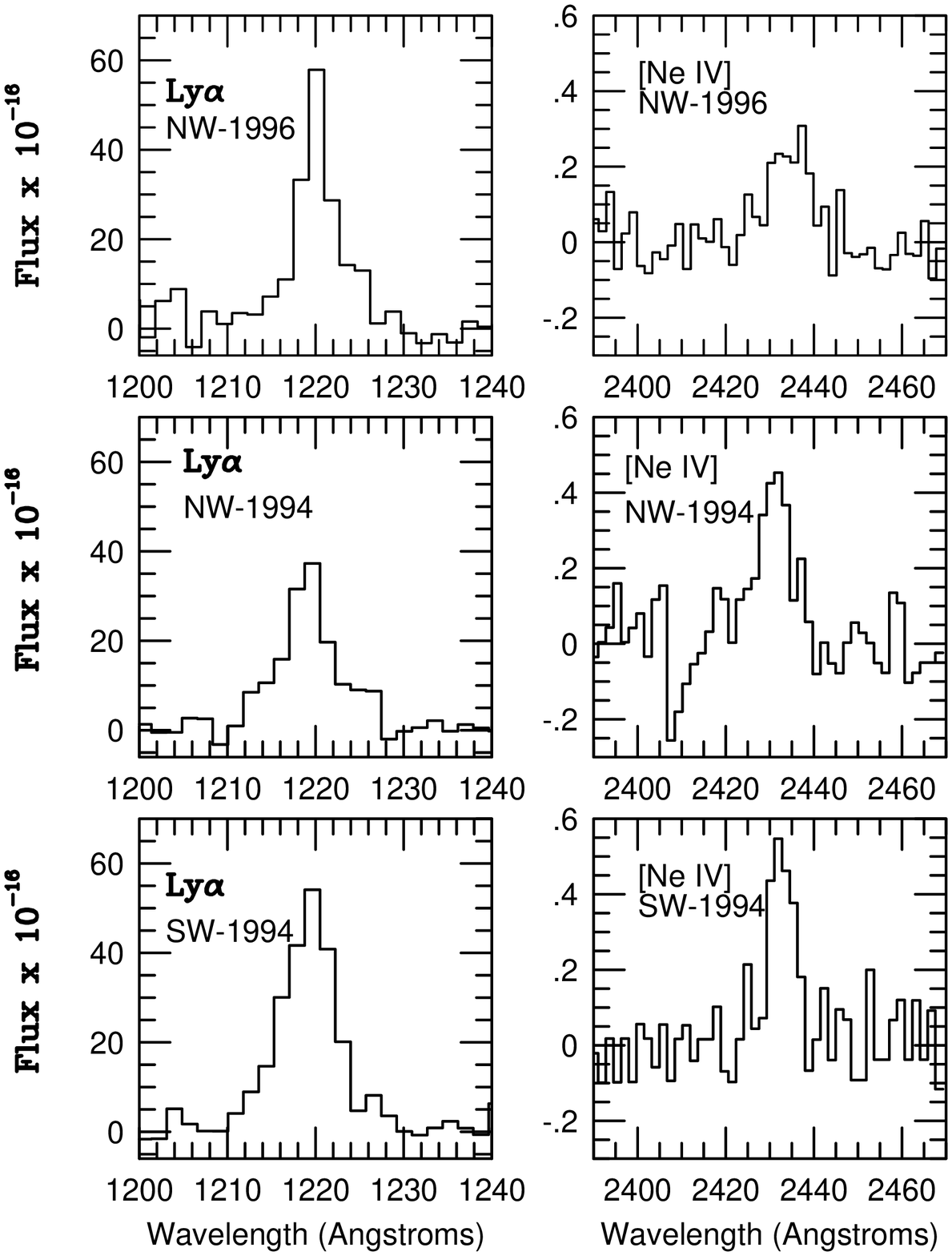]{Expanded spectra near the Ly${\alpha}$ (left)
and [Ne\,{\sc iv}] lines (right).  The top spectra show the lines in
the 1996 NW pointing while the middle spectra show the lines in the
1994 NW pointing.  The bottom spectra show the corresponding lines in
the 1994 SW pointing.}
\scalebox{0.5}{\includegraphics{sn78k_fig6.ps}}
\label{HST_spec_off}
\end{figure*}

The ``on-source'' exposure time is therefore a function of one's
definition of ``on-source.''  The ``true'' on-source observation is
the red spectrum; no observation in the blue was obtained at that
location.  The observations obtained NW of SN1978K lie $\sim$1''.3
from the supernova; the observations obtained to the SW lie
$\sim$2''.3 away.  These distances are approximately 4.3 and 7.6 times
the size of the FOS aperture or approximately 6.5 and 11.5 times the
70\% encircled energy radii of the COSTAR-corrected {\it HST} optics.

The data were calibrated using the standard FOS pipeline.  We checked
the wavelength calibration at the end points of the spectra and
verified the flux calibration using FOS tasks in STSDAS\footnote{The
Space Telescope Science Data Analysis System is distributed by the
Space Telescope Science Institute.}.  We carefully examined each
spectrum for evidence of noisy or dead pixels.  The red spectrum
centered on the SN had a noisy pixel in the middle of the band; we
clipped out the affected pixels.  None of the candidate lines
correspond to noisy pixels as listed in the August 1994 noisy pixel
list \citep{HST95}.  All of the candidate lines were
visible in the raw and calibrated data.  We then co-added the data for
a given pointing and epoch.  The spectra are shown in
Figures~\ref{HST_spec_on} and~\ref{HST_spec_off}.

All line fluxes were corrected for reddening using two different
values.  The first value of $E_{\rm B-V} = 0.01$, comes from
\cite{BH84} while the second value, $\sim0.31$, comes from
\cite[Paper I]{R+93}.  These values were converted to A$_{\lambda}$
using the reddening curve in \cite{Ma90}.  The first value increases
line fluxes by a factor of $\sim1.06$ at $\lambda$ $\sim$2800{\AA}
while the second increases the fluxes by a factor of $\sim5.5$.  The
second value would seem more likely, given the environment immediately
after a supernova explosion.  However, the optical spectrum (to be
presented shortly) implies that the extinction has decreased to
approximately zero.

\subsection{Optical Light Curve}

\begin{figure*}
\caption[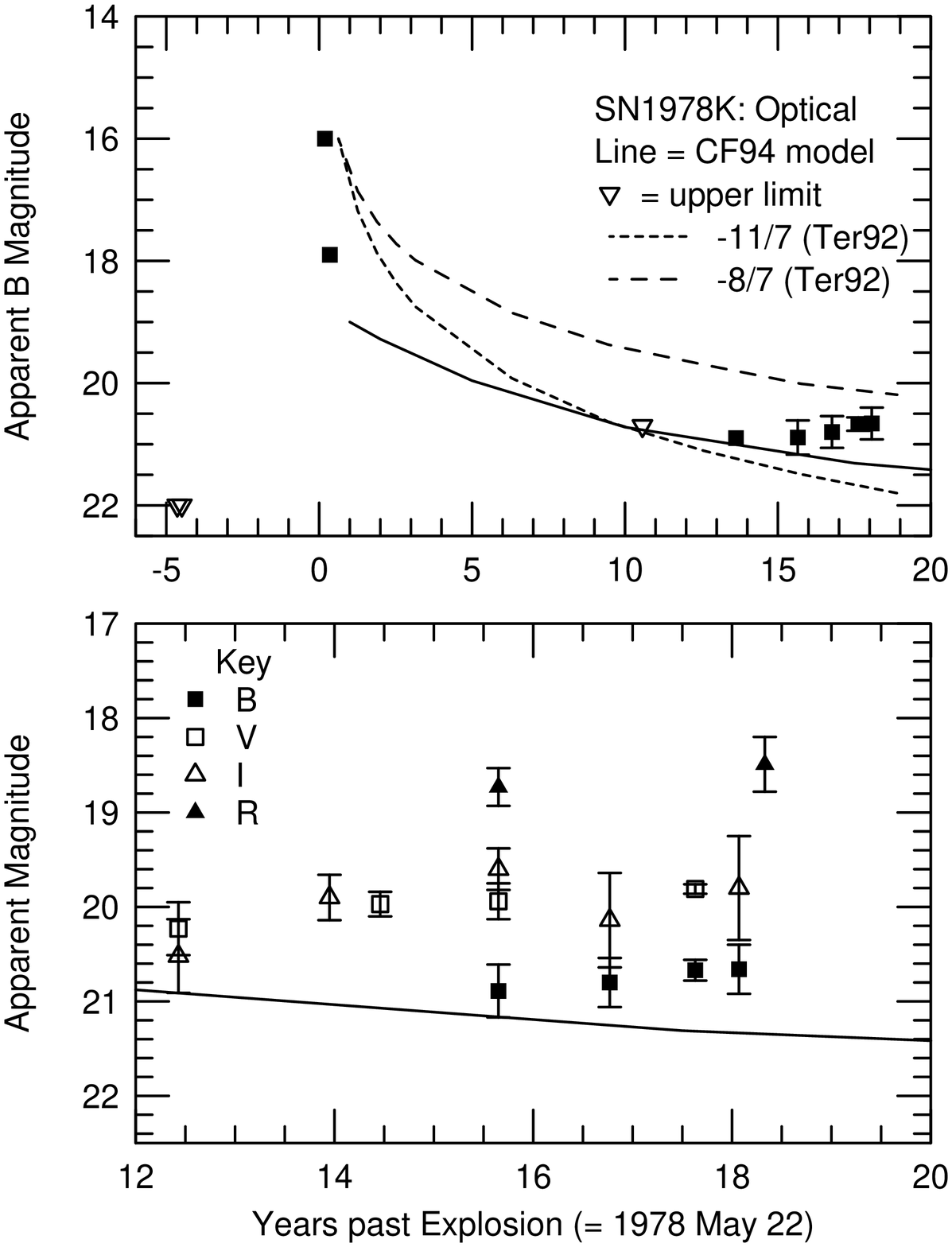]{The optical light curve.  The top panel shows
the complete B band light curve; the bottom panel expands the recent
history and includes the V, R, and I points as well as the {\it HST}
pseudo-B and -V points.}\label{optlc}
\scalebox{0.6}{\includegraphics{sn78k_fig7.ps}}
\end{figure*}

The optical light curve originates from two sources:  ground-based
optical photometry and WFPC2 snapshots using {\it HST}.

To obtain accurate coordinates and offset stars for SN~1978K in the
{\it HST} reference frame, $B$ and $V$ band WFPC2 snapshots of the
SN~1978K region were obtained.  The observations occurred on 1996
January 3 with exposure times of 60~sec for both the F555W ($\sim V$)
filter and the F439W ($\sim B$) filter.  We extracted the counts for
SN~1978K as well as the counts for stars `b' ($\sim 66''$ E of
SN~1978K), `c' ($\sim42''$ W), and `d' ($\sim44''$ NW) (q.v., Paper~I
for a finding chart) using apertures $3''$ in diameter.  We adjusted
the resulting magnitudes for the gain, charge transfer efficiency,
contamination corrections, and added the zero offset to place the
magnitudes relative to Vega.  These corrections are all described in
the \cite{HST95}{\it HST} Data Handbook (version 2.0). For the $V$
band, the magnitudes of stars b, c, and d differed from their
published values (Paper~I) by a mean value of +0.759; for the $B$
band, the mean value was +1.396.  Correcting the SN~1978K magnitudes
by these offsets gave the $B$ and $V$ magnitudes for SN~1978K:
20.67$\pm$0.11 and 19.81$\pm$0.05, respectively.

The supernova has also been monitored intermittently on our behalf by
various observers. The resulting magnitudes, determined from
differential photometry relative to the sequence established in
Paper~I, are collated in Table~\ref{optmags} and shown in
Figure~\ref{optlc}. The data are not conclusive; an increase in
brightness of $0.2-0.3$~mag in all filters is suggested, but the
possibility that there has been no change in brightness of SN~1978K
between 1990 and 1996 cannot be ruled out.

\subsection{Optical Spectrum}

\begin{figure*}
\caption[sn78k_fig8.ps]{Optical spectrum of SN~1978K obtained on
1996~October~8 with the RGO Spectrograph on the Anglo-Australian
Telescope. Major lines are identified in
Table~\protect{\ref{aatlines}}. No dereddening or redshift corrections
have been applied.}
\rotatebox{-90}{\scalebox{0.6}{\includegraphics{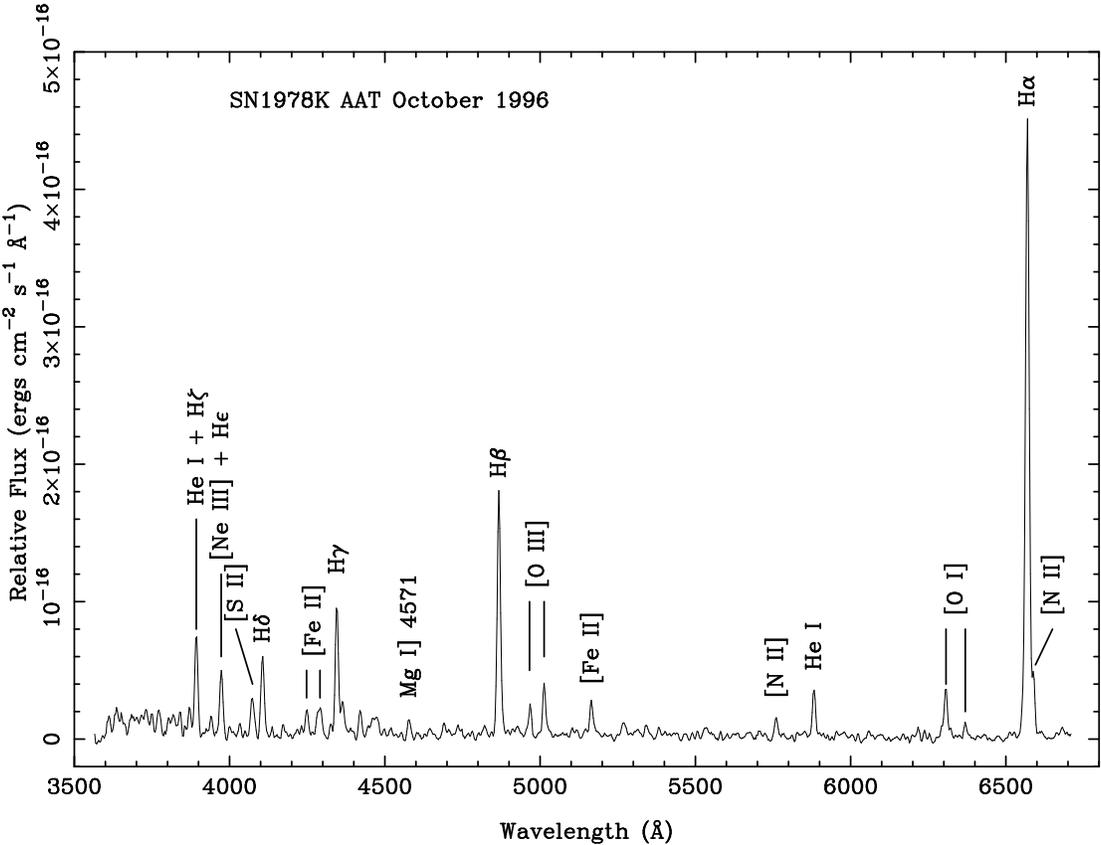}}}
\label{optspec}
\end{figure*}

Spectrophotometry of SN~1978K in the wavelength range $3566-6715$~\AA\
was obtained on 1996~Oct~8 UT with the 3.9~m Anglo-Australian
Telescope. The RGO Spectrograph was used with the 25~cm camera, a
300~lines~mm$^{-1}$ grating, a slit length of 77'', and a Tektronix
1K CCD.  Three consecutive exposures of 1000~sec each were obtained on
the supernova, followed by a short exposure on the flux standard EG21
\citep{Ham94}. The three exposures were reduced
individually within {\sc iraf}\footnote{IRAF is distributed by the
National Optical Astronomy Observatories, which are operated by the
Association of Universities for Research in Astronomy, Inc., under
cooperative agreement with the National Science Foundation.}, and then
the line fluxes were measured separately for each spectrum to permit
the flux errors to be calculated. The final resolution obtained after
light smoothing is 12~\AA. The result of co-adding all 3~spectra is
shown in Figure~\ref{optspec}, and the line fluxes (relative to
H$\beta=100$ and uncorrected for reddening) and proposed
identifications are listed in Table~\ref{aatlines}.

\subsection{Radio Observations}

\begin{figure*}
\caption[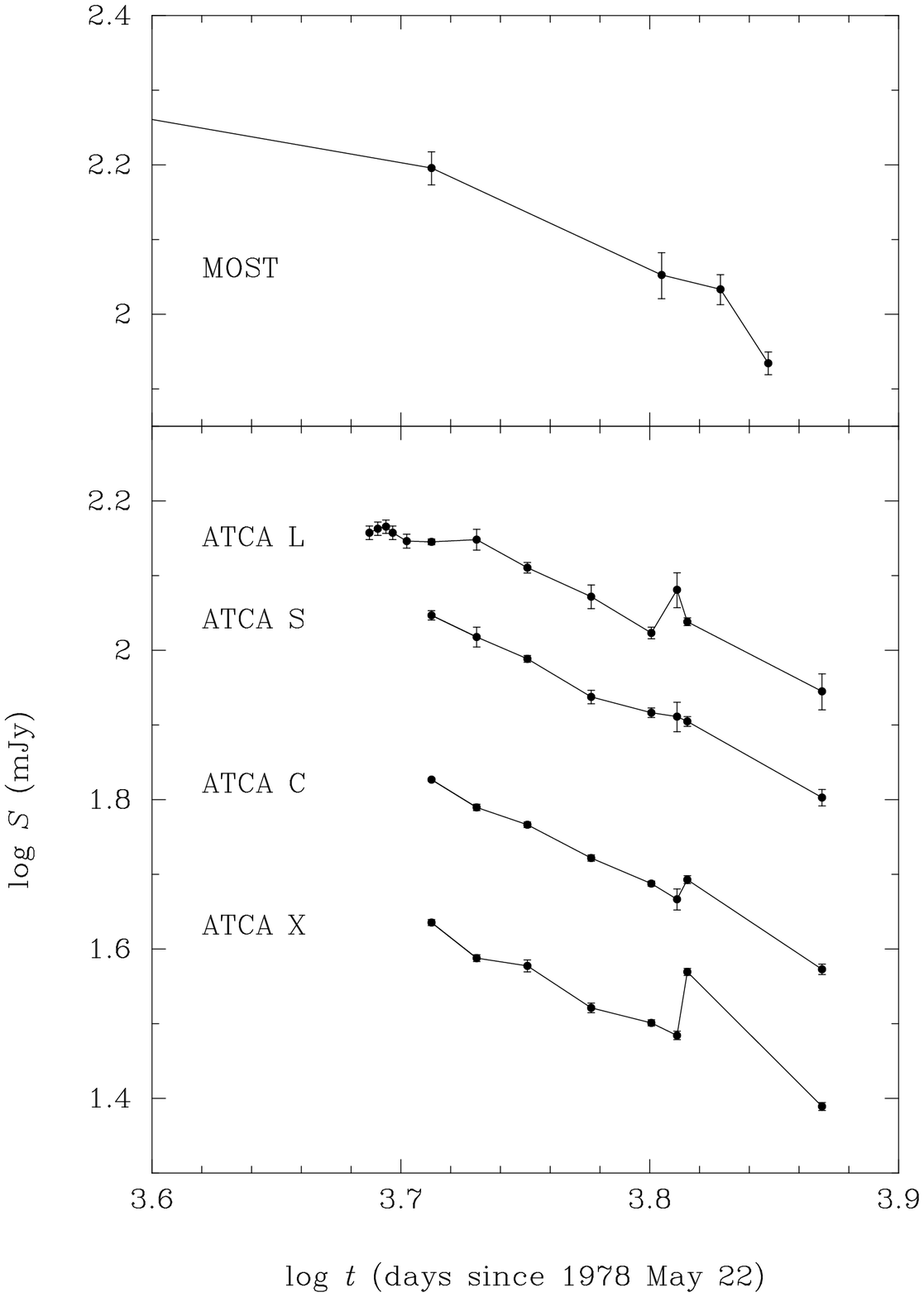]{The post-1990 radio light curve for SN~1978K,
incorporating data from this paper, Paper~I, and from
\cite[MWP]{MWP}. The ATCA and MOST data are plotted on separate
panels, but the time and flux scales are the same for both. The time
axis is expressed as log of the number of days elapsed since 1978 May
22 for consistency with \cite[MWP]{MWP}.}
\scalebox{0.6}{\includegraphics{sn78k_fig9.ps}}
\label{rlc}
\end{figure*}

SN~1978K has been monitored at irregular intervals using the
6~antennas of the Australia Telescope Compact Array (ATCA) since its
discovery.  Using the multi-frequency agility of the ATCA,
observations with a 128~MHz bandwidth are taken over an 4-12~hour
period, with rapid switching between simultaneous measurements at 1380
and 2370~MHz, or at 4790 and 8640~MHz. Beginning in August~1998, the
central frequency of the S-band observations was changed to 2496~MHz
in an effort to avoid encroaching interference. The observations were
made with SN~1978K displaced slightly from the phase-tracking center
to avoid the most common system-related artefacts. The continuum data
over all 15 baselines was examined interactively using {\sc aips} to
remove interference, then calibrated using repeated measurements of
the secondary flux calibrator, PKS B0252-712.  The flux scale was set
by using observations of PKS B1934-638, and the revised calibration of
\cite{Rey}.

\begin{figure*}
\caption[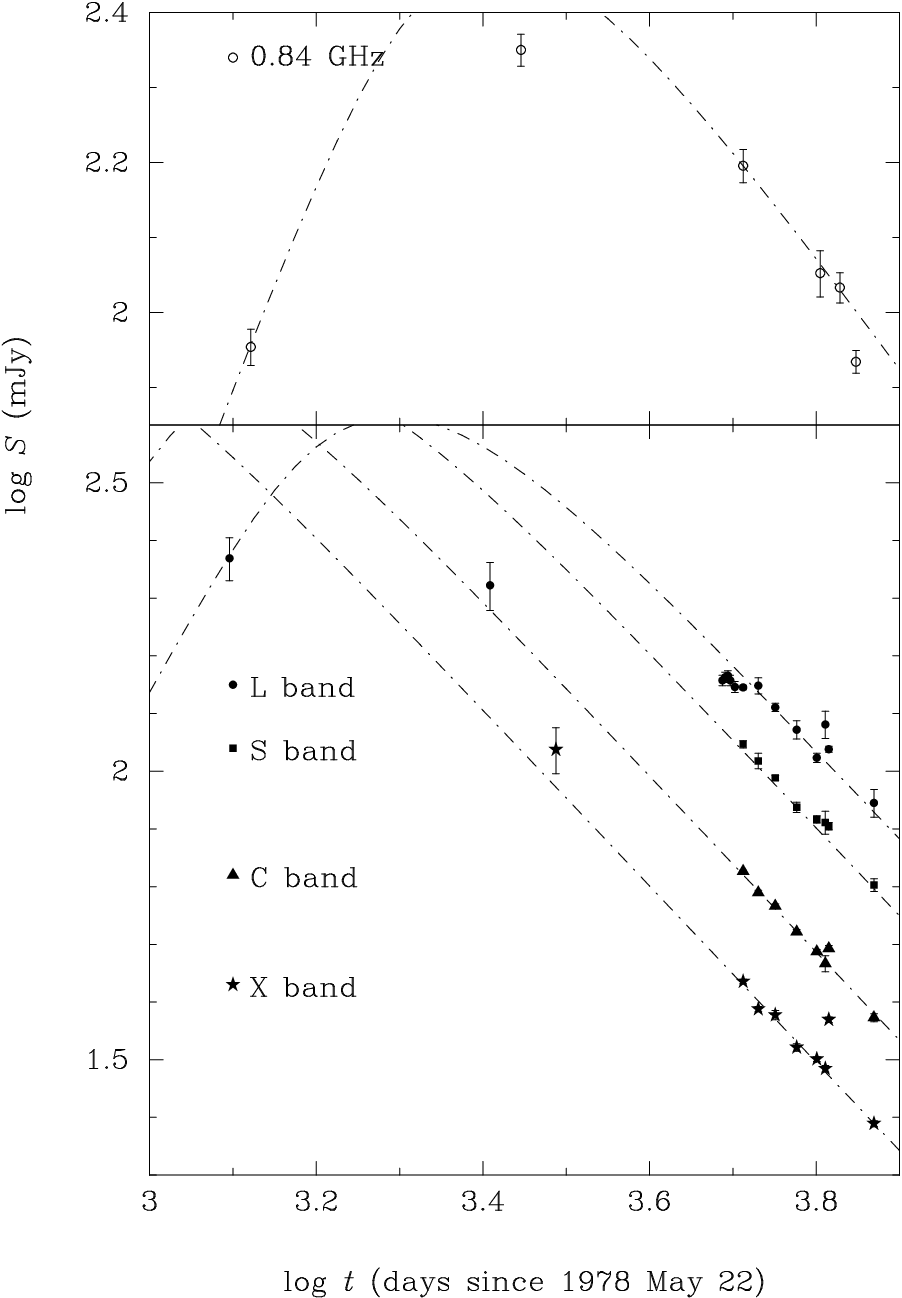]{Plots of the full 17-year dataset of ATCA and MOST
observations of SN 1978K, together with the pre-discovery observations
by \cite{Peter94}. Our best overall fit, using the modified
version of the \cite{W86,W90} model described in \cite[MWP]{MWP},
for the five frequency bands is shown by the dot-dashed lines. The
parameters for this fit (cf. Table 10) are as follows: $K_{1}1 =
3.1\times10^{7}$~mJy, $\alpha = -0.76$, $\beta = -1.53$, $K_{2} =
6\times10^{4}$, $K_{3}=1.05\times10^{11}$, and $K_{4} =
1\times10^{-2}$.}
\scalebox{0.6}{\includegraphics{sn78k_fig10.ps}}
\label{rlcmod}
\end{figure*}

Deconvolved images of SN~1978K show the source still to be unresolved
at all frequencies. Since SN~1978K is the only significant source of
continuum emission in the field (Fig.~5 of Paper~I), it was decided to
measure the fluxes directly from plots of the calibrated visibility
amplitudes, binned into $10-20$ time periods, to give a truer
reflection of the errors, which are dominated more by atmospheric
conditions than by the model fitting to the deconvolved images. The
results from eight epochs are compiled in Table~\ref{atcafluxes},
together with three observations at 0.843 GHz using the Molonglo
Observatory Synthesis Telescope (MOST).  The post-1990 radio light
curve, including data from Paper I and \cite[MWP]{MWP}, is
presented in Figure~\ref{rlc}.

\section{Analyses, by Band}

\subsection{X-ray Light Curve Modeling}

SPC96 presented the first X-ray light curve for SN1978K and showed
that the light curve was consistent with zero slope.  The updated
X-ray light curve (Figure~\ref{HRI_LC}) should decay as t$^{\alpha}$
(e.g., \cite{CF94}).  The formal fitted
value of the exponent to all of the data (flux versus days past
maximum ) is ${\alpha} = (-3.71\pm6.52) \times 10^{-5}$ (90\%
confidence range) if all of the data points are included in the fit.
This fit is illustrated in the figure as the solid line. The dashed
lines show the $\pm$90\% range of the slope. Note that the range is
consistent with zero. As a check, we fit only the HRI data points,
obtaining a value for ${\alpha} = +(5.60\pm11.72) \times 10^{-5}$.
This slope also includes zero and covers approximately the same range.
Additional observations must be obtained to ascertain whether SN~1978K
has started its slide into obscurity or if the last data point is a
fluctuation.  Given the scatter of the other data points, a
fluctuation provides the most likely explanation.  Fluctuations in the
X-ray light curve with amplitudes of $\sim$20\% are expected if the
ejecta are inhomogeneous (e.g., \cite{Cid96}).

\cite[FLC96]{FLC96} discussed several regimes for the decay of X-rays
from SN~1993J.  Those regimes were (i) the optically thick shock case;
(ii) the adiabatic case; and (iii) the radiative case.  The time
behavior from each differed.  For the optically thick case, $L \propto
T^{0.16} t^{(3 - 2s)(n - 3)/(n - s)}$ where $T = $temperature, $t =
$time, $n = $power law index for the density profile of the ejecta,
and $s = $power law index for the density profile of the circumstellar
matter.  The power law is an approximation to the supernova density
profile; typical values from models lie between 7 and 12.  The CF94
model requires $n \lesssim 9$ for an adiabatic shock; for $n \gtrsim
8$, radiative cooling becomes important.  A steady stellar wind will
create a circumstellar medium with $s = 2$, so that value is usually
adopted.  But FLC96 showed that $1.5 \lesssim s \lesssim 1.7$
described the SN~1993J X-ray behavior more accurately.

For the adiabatic case, $L \propto t^{-[(2s - 3)n - 5s + 6]/(n - s)}$,
while for the radiative case, $L \propto t^{-(15 - 6s + sn -2n) / (n -
s)}$.  If we assume the slope is precisely zero, then we can explore
the possible values for $n$ and $s$.  A value of $s = 2$ is usually
adopted because it describes the circumstellar density distribution
from a steady wind. A value of $\sim1.5$ for $s$ is supported by the
SN~1993J data (FLC96), however, subsequent analysis shows $s = 2$ is
correct \citep{FB98}.  While $s = 2$ has physical significance, we
will also explore the effect of $s = 1.5$. For the optically thick
case, either low $n$ ($\sim4$) or $s = 1.5$ plus any $n$ yields a
solution.  For the adiabatic case, if $s = 2$, $n$ must again be low
($\sim4$), while $s = 1.5$ leads to no solution.  For the radiative
case, if $s = 2$ leads to no solution, while if $s = 1.5$, then $n =
12$.

Using just the light curve, we are unable to obtain sufficient
information to establish the emission behavior uniquely because either
the optically thick shock or the radiative cases fit the data equally
well.  We only establish a range for $n$ ($\sim$4-12).  Further, these
two models use $s = 2$ and $s = 1.5$.  By comparison, hydrodynamic
models establish a power-law distribution with $n~\sim8-20$ within a
day or two of the explosion \citep{Chev90}.  Only additional
observations across a wider time span will yield sufficient
information to break the degeneracy and allow the model to be directly
tested.

\subsection{Analysis of the {\it ASCA} X-ray Spectrum}\label{asca_analy}

Each epoch's spectrum was fit using simple, absorbed models
(bremsstrahlung, power law, and emission from a hot, diffuse gas)
(Table~\ref{asca_fits}).  Within the errors, the model fits were
identical for the two epochs regardless of the model.
Figure~\ref{asca_spec}(a) shows the 1995 November spectrum and
Figure~\ref{asca_spec}(b) shows the fitted contours both for the hot
gas model.  We present the spectral fit using the model with a thawed
abundance to show the range of the parameters and the degeneracy of
the model.  The fitted 1993 spectrum in \cite{Pet94b} fixed the
abundance at 1.0; when we use the same fixed abundance for the 1995
spectrum, we obtain the same fitted temperature as did Petre et al.
The corresponding figure for the first epoch appears nearly identical
\citep{Pet94b}.  No one model provided a significantly better fit than
the others.  We integrated the model spectra across several bands
(0.5-2.0, 2.0-3.0, 3.0-4.0, 4.0-5.0, and 5.0-9.0 keV) to obtain fluxes
in each band for each epoch.  The fluxes are also consistent with each
other within the errors.  No lines were detected in either epoch.  The
unabsorbed 0.5-2.0 keV flux is $\sim$1.3$\times$10$^{-12}$ ergs
s$^{-1}$ cm$^{-2}$ which corresponds to a luminosity of
$\sim$3$\times$10$^{39}$ ergs s$^{-1}$.  This flux lies within 10-20\%
of the mean flux obtained with {\it ROSAT}, depending upon the adopted
model for the X-ray spectra.

The luminosities of the {\it ASCA} (1-10 keV) and {\it ROSAT} (0.1-2.4
keV) bands provide an estimate of the shock temperature
\citep[FLC96]{FLC96} via $L_{\rm ASCA} / L_{\rm ROSAT} \sim
2.4e^{-{7.9{\times}10^7}/T_e}.$ Our luminosities yield T$_e$
$\sim$6${\times}$10$^6$ K.  This is the temperature of the reverse
shock where soft X-rays are expected to be produced \citep{CF94}.

\subsection{{\it HST} FOS Spectrum}

As expected for a nebular object, we do not see a continuum in the
{\it HST} spectra.  The pointing history makes the UV spectra
(Figures~\ref{HST_spec_on} and \ref{HST_spec_off}) difficult to
interpret.  At the distance of NGC 1313, 1'' is $\sim$22 pc, so we are
sampling relatively widely-spaced regions with the FOS spectra.

The Mg\,{\sc ii} 2800 {\AA} line is a good starting point because the
line is the easiest to interpret.  It is visible in the on-source
spectrum and not to the NW.  This is a clear detection of SN1978K.
The Mg\,{\sc ii} $\sim$2800 {\AA} is a doublet with components at 2795
and 2803 {\AA}.  We fit the lines with double gaussians and
constrained the FWHM to be identical for the two components.  The
results of the fits are shown in Table~\ref{FOS_lines}.  The fitted
FWHM corresponds to the instrumental resolution, so the lines are
unresolved.  We derive velocities for the gas of 632 and
640~km~s$^{-1}$, respectively.  With errors on the line centers from
the gaussian fits of order 10-15 km s$^{-1}$, the two line velocities
are identical.  The redshift for SN~1978K, taking into account the
rotation velocity of NGC~1313 at the position of the supernova, is
439~km~s$^{-1}$ (Paper~I); the lines have intrinsic velocities of
$\sim200$~km~s$^{-1}$.

The Ly${\alpha}$ and [Ne~IV] lines are detected in the NW and SW
pointings.  We need not discuss Ly${\alpha}$ because it is emitted by
nearly every source in the sky.  The [Ne~IV] line is not observed in
the on-source pointing.  It must arise from a location farther from
SN1978K than the projected aperture size.  The FOS aperture is
$\sim$0''.15 in radius, which corresponds to $\sim$3.3 pc at NGC 1313.
The NW and SW pointings lie at projected spatial distances of $\sim$28
and $\sim$50 pc from SN1978K.  While the gas velocities of the [Ne~IV]
lines are larger, by a factor of two, than the velocities measured
from the optical and Mg II (UV) emission lines, the projected
distances are far larger than the distance any ejecta of SN1978K could
cover in the $\sim$20 years since its detonation.  This is
particularly critical given that the supernova is supposed to be
evolving in a dense circumstellar medium which means that the ejecta
are rapidly decelerated (e.g., \cite{Ter92}).  If we assume that the
X-ray flux has illuminated the surrounding volume for the entire
lifetime of SN1978K, the volume's radius is $\sim$20 light years or
$\sim$6 pc.  The [Ne~IV] is produced outside of this volume and must
either be located in the pre-shock medium or is unrelated to SN1978K.
If the [Ne~IV] matter moved at its measured velocity, it would cover
the $\sim$30-50 pc in $\sim$42-70 Kyr which is entirely consistent
with durations of mass loss phases in massive stars (e.g.,
\cite{MNG98}).

The UV line fluxes to the NW, the only position for which we
have spectra at both observation epochs, are constant within the
errors across the two-year gap. This is expected if these lines
originate outside of the shocked material.

We also include in Table~\ref{FOS_UL} upper limits on ultraviolet
lines that are typically found in other SNR and H II regions and have
been predicted by models (e.g., \cite{Ter92,CF94}) These upper limits
were estimated by fitting a gaussian to noise spikes at the location
of the expected emission line and were dereddened using the adopted
two values above.

\subsection{Optical Light Curve and Optical Spectrum}\label{aat_spec}

The optical light curve (Figure~\ref{optlc}) shows the predicted
behavior of the interaction model of \cite{CF94} assuming an
explosion date near 1978 June 1.  If the slight rise in the optical
magnitudes continues, it will increasingly be at odds with the
prediction (i.e., decline) of this model.  The increase can not be
attributed to a simple increase in emitting area because the necessary
expansion velocity, $\sim$2000 km s$^{-1}$ is not supported by the
observations.  Figure~\ref{optlc} also shows the predicted -11/7
(leading shock) and -8/7 (material behind shock) decays from
\cite{Ter92} (their equations 5 and 6).  The two curves have been
arbitrarily normalized to magnitude 16 at 230 days (1 e-fold time).
The arbitrary normalization was chosen to show that the -8/7 curve
decays too quickly.  If the increase in brightness continues, both
curves increasingly fail.

The optical spectrum is presented in Figure~\ref{optspec}.  The most
significant change since the 1990 and 1992 spectra presented in
Paper~I is that the Balmer line ratios indicate that the line of sight
extinction towards SN~1978K has dropped from $A_B \sim 2$~mag to
virtually zero. Despite using a wide slit in moderate seeing
($1\farcs5$), it is possible that differential refraction could cause
us to lose more of the red light than the blue (the slit was oriented
N-S and not at the parallactic angle). Indeed, with a drop in the
extinction of 2~magnitudes, we might have expected SN~1978K to have
brightened optically in that time, but the optical photometry
(Table~\ref{optmags}) suggests little or no increase in brightness.
We conclude that radiative transfer effects have altered the
decrement.  Table~\ref{Balm_decr} lists the Balmer line strengths for
the 1990, 1992 \citep{R+93}, and 1996 observations, along with values
for Case A and Case B recombination assuming T = 5000 K (Case A) and T
= 10$^4$ K and N$_{\rm e}$ = 10$^6$ cm$^{-3}$ (Case B) \citep{Ost89}.
For Case A at T = 10$^{4}$ K, the values are nearly identical to Case
B.

Extended emission is visible in the spatial direction.  About 8''.5 N,
emission is present in H$\alpha$, H$\beta$, and [O~III]~3727{AA}.
This matches very well with a nearby, compact H II region just visible
in the plate in \cite{R+93}.

The [O\,{\sc iii}] lines provide an estimate of the temperature
([I(4959)+I(5007)]/I(4363)) of $\sim$1-1.5$\times$10$^4$ K
\citep{CKA86} for n$_{\rm e}$ $\sim$10$^6$ cm$^{-3}$ if we assume all
of the 4363{\AA} emission is [O~III].  The estimated temperature
changes very little if n$_{\rm e}$ varies from 10$^5$ to 10$^7$
cm$^{-3}$.

None of the lines is resolved which implies that the shock velocity is
low; the instrumental resolution is of the order of 10-12{\AA} which
corresponds to $\sim$500-600 km s$^{-1}$.  These values are similar to
the velocities in the {\it HST} spectrum.  

\subsection{Radio Light Curve Modeling}

With the exception of the 1996 observations, SN~1978K continues to
show a fairly uniform decline in radio luminosity, over the frequency
range $1.4-8.6$~GHz, given by

\begin{equation}
S \propto (t-t_{0})^{\beta}
\end{equation}

\noindent
with $\beta=-1.53\pm0.13$ and $(t-t_{0})$, the number of days since
the explosion ($t_{0}$ assumed to be 1978 May~22 UT as suggested by
\cite{MWP}). 

Several of the radio points deviate from the fitted curve.  Although
the MOST coverage at 0.843~GHz spans a longer time baseline,
observations are less frequent, but there are indications that the
rate of decline may be accelerating at the lower frequencies.
Observations at the L-band near day 5000 (log t = 3.7) fall below
the fitted curve.

Near the beginning of 1996, the supernova remnant appears to have
undergone a brief flare-up in brightness, which shows up first in the
L-band, then a couple of months later at the higher frequencies, but
has since returned to close to its normal rate of decline.
Unfortunately, only one observation at another bandpass was obtained
near this flare: an optical B magnitude, which shows no evidence of a
change in the optical light.

The time and spectral evolution of Type~II radio supernova has been
successfully modeled by Weiler et al. \citep{W86,W90} using a
variation on the minishell model of \cite{Chev82}. \cite[MWP]{MWP}
attempted to apply this model to the ATCA and MOST data up to 1992
July, but were unable to fit the data at all five frequencies using
just a single value for the radio spectral index $\alpha$. In addition
to being attenuated by a uniform absorbing medium of varying optical
depth $\tau$, and a clumpy external absorbing medium of varying
optical depth $\tau^{\prime}$, they needed to invoke an extra (time
invariant) absorption, such as might be produced by an H\,{\sc
ii}~region along the line of sight to SN~1978K, in order to account
for the fact that SN~1978K appears sub-luminous at late epochs in the
low frequency regime probed by the MOST. Their new model has the form

\begin{equation}
S = K_{1} \left( \frac{\nu}{5~{\rm GHz}} \right)^{\alpha}
    \left( \frac{t-t_{0}}{1~{\rm day}} \right)^{\beta}
    e^{-(\tau + {\tau}^{\prime\prime})} \left( \frac{1 - e^{-\tau^{\prime}}}
    {\tau^{\prime}} \right) ~{\rm mJy} ,
\label{eq:fullmod}
\end{equation}

\noindent
where

\begin{equation}
\tau = K_{2} \left( \frac{\nu}{5~{\rm GHz}} \right)^{-2.1}
       \left( \frac{t-t_{0}}{1~{\rm day}} \right)^{\delta} ,
\label{eq:tau}
\end{equation}

\begin{equation}
\tau^{\prime} = K_{3} \left( \frac{\nu}{5~{\rm GHz}} \right)^{-2.1}
       \left( \frac{t-t_{0}}{1~{\rm day}} \right)^{\delta^{\prime}} ,
\label{eq:tau2}
\end{equation}

\noindent
and

\begin{equation}
\tau^{\prime\prime} = K_{4} \left( \frac{\nu}{5~{\rm GHz}} \right)^{-2.1} .
\label{eq:tau3}
\end{equation}

\noindent
$K_{1}$ and $K_{2}$ are the unabsorbed flux density and the optical
depth in the uniform absorbing medium at $\nu=5$~GHz one day after the
explosion, respectively; $K_{3}$ is the optical depth in a clumpy,
non-uniform medium at the same epoch; $K_{4}$ is the (time
independent) optical depth towards SN~1978K at 5~GHz due to thermal
ionised hydrogen; and $\delta$ and $\delta^{\prime}$ are related to
the combination of the frequency and time dependence by $\delta \equiv
\alpha - \beta - 3$ and $\delta^{\prime} \equiv 5\delta / 3$
\citep{MWP,W90}.

We have attempted to apply the model described by
eqtn.~\ref{eq:fullmod} to the full ATCA $+$ MOST dataset (now covering
twice the time baseline available to \cite[MWP]{MWP}). We also
include here, for the very first time, three pre-discovery
observations taken from \cite{Peter94}: a 1420 MHz observation with
the Two-Element Synthesis Telescope (TEST) in 1981; a 1415 MHz
observation with the Fleurs Synthesis Telescope in 1985; and a 8420
MHz observation with the Tidbinbilla 64m Deep Space Network antenna in
1986. These data allow us to solve simultaneously for all the free
parameters in eqtn.~\ref{eq:fullmod}, including the $\tau^{\prime}$
term which influences the early evolution of the radio supernova, but
which was neglected by \cite[MWP]{MWP} owing to insufficient data at
early epochs.

Our best fit model is shown in Figure~\ref{rlcmod}, which has
$K_{1}=3.1\times10^{7}$~mJy, $\alpha=-0.76$, $\beta=-1.53$,
$K_{2}=6\times10^{4}$, $K_{3}=1.05\times10^{11}$, and
$K_{4}=1\times10^{-2}$. Despite the large number of free parameters in
the model, the difficulty in simultaneously fitting the data at all
five frequencies in both the early and late phases of evolution is
well illustrated in this figure. No combination of parameters can make
the L-band model fit both of the early \cite{Peter94} points, and fit
the data at late epochs. By balancing the values of mainly $\beta$,
$K_1$, and $K_{3}$, it is possible to arrive at a solution that fits
both the TEST and earliest MOST observations (Figure~\ref{rlcmod}),
but which still predicts too much flux in these bands at, or just past
the maximum. We were able to identify a solution with $\beta=-1.90$
which fits all the MOST points except the last, but then the fit to
the very latest points at all other frequencies is also poor. We have
explored the range of parameter space which allows acceptable fits to
the bulk of the data, and tabulate the allowable ranges in Table 10.

We show in this same table the parameters arrived at by
\cite[MWP]{MWP}, as well as comparable models for other Type~II SNe
from the literature. We find best agreement with \cite[MWP]{MWP} on
the spectral index $\alpha$, and on the absorption due to the
foreground H\,{\sc ii}~region $K_{4}$.  The most significant
difference arises from our post-1992 data, which clearly requires a
much steeper rate of decline than that suggested by \cite[MWP]{MWP}.
Consequently, we also require an initial flux density $K_1$ which is
as much as 3~orders of magnitude larger, coupled with a similar
reduction in initial optical depth $K_2$ from uniform obscuration, if
the model is to come close to accounting for the observed evolution
over a full decade in frequency and more than 15 years in time. Our
fit to the new pre-discovery data points yields a value for the
external, non-uniform absorption $K_{3}$ which is within the upper
bounds established by \cite[MWP]{MWP}.

We further conclude that on the basis of the range of allowable model
fits, SN~1978K reached its peak 5 GHz luminosity some 240--300 days
later than claimed by \cite[MWP]{MWP} (i.e., 940--1000 days after
the explosion), and that it was at least 50\% more luminous
(350--830~mJy). This then brings SN~1978K up into the same peak
luminosity range as the highly-luminous Type~IIn supernovae SN~1986J
and SN~1988Z (Table~\ref{mwpar}).  Furthermore, SN~1978K is now
showing the same type of very steep decline. With the benefit of the
new radio observations at late epochs, together with the inclusion of
extra pre-discovery data, we must refute the statement of
\cite[MWP]{MWP} that ``...SN~1978K was almost certainly a fairly
normal Type~II supernova'', and instead restore it to the class of
radio and X-ray luminous Type~II supernovae exemplified by SN~1986J
and SN~1988Z.

\section{Discussion}

A detailed picture is emerging for the X-ray luminous supernovae,
particularly the SN IIn variety.  The supernova explodes into a very
dense ($\sim$10$^{6-8}$ cm$^{-3}$) circumstellar medium.  In such a
medium, the evolution of the supernova is accelerated, so the evolving
remnant goes directly from the explosive phase to the radiative phase.
The pioneering theoretical studies for supernovae exploding in dense
media were carried out over the past 25 years
\citep{Chev74,Wheel80,Shu80,Ter92}.  Terlevich et al. coined the term
``compact SNR'' because the bulk of the original kinetic energy
($\sim$10$^{51}$ ergs) is radiated while the supernova is
$\sim$10$^{16-17}$ cm in size.  The object then radiates large
quantities of energy in the X-ray, ultraviolet, and radio bands.  The
velocities of the emission lines are low ($\lesssim$10$^3$ km
s$^{-1}$) representing the extreme deceleration that has occurred.

\begin{figure*}
\caption[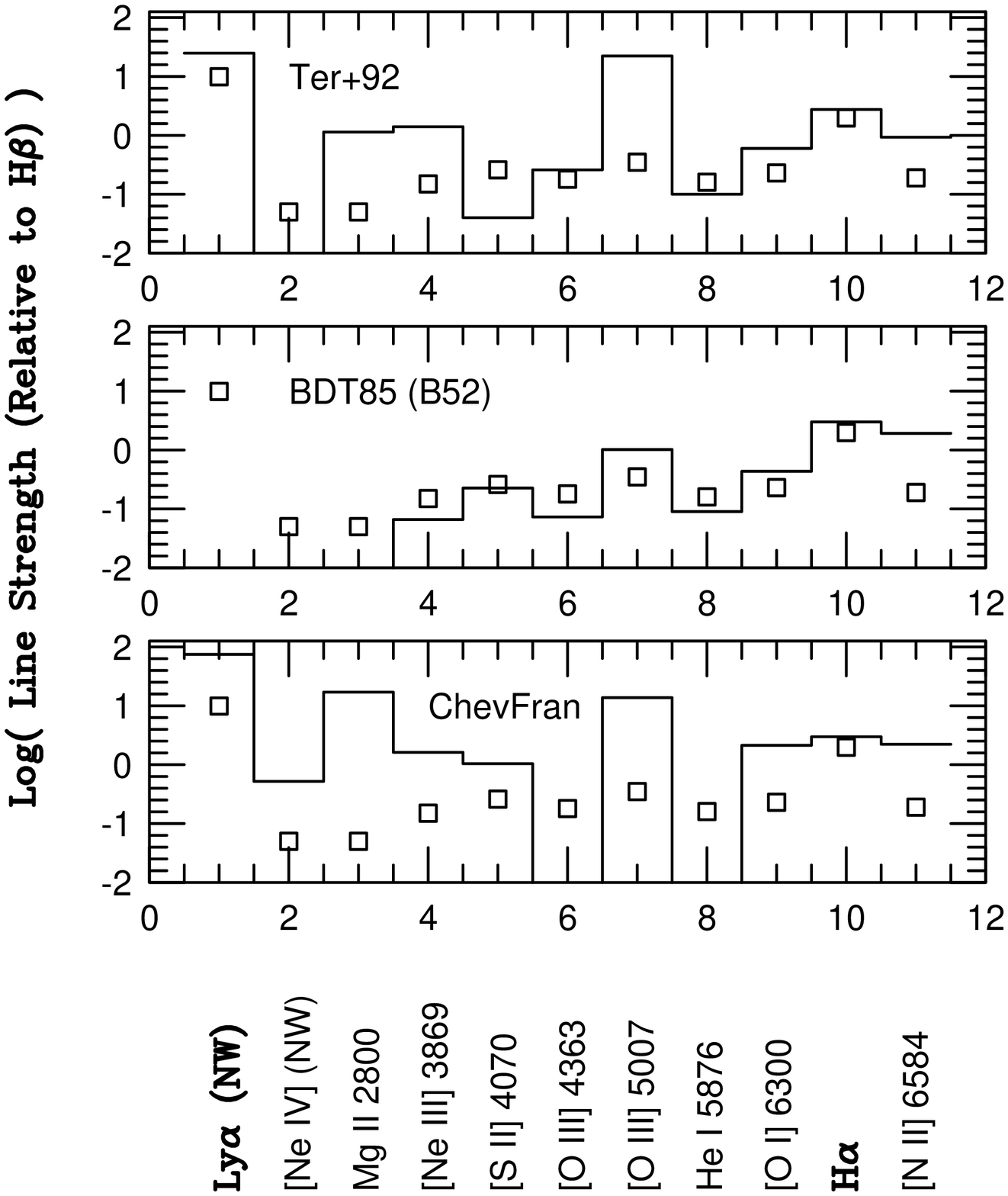]{A comparison of the UV-optical line strengths
of SN1978K with published models.  The two panels are: (top) Ter+92:
the compact SNR model for the Narrow Line Region of AGN by
\cite{Ter92}; (middle) the radiative shock model B52 \citep{BDT85};
and (bottom) ChevFran: the 17.5 year model for interacting supernovae
from \cite{CF94}.  The lines are identified below the bottom plot.
The open squares are the observed line strengths; the solid lines
connect the model predictions.  Any model which did not predict a line
strength was assigned an arbitrary value of log( strength ) = -2.0.}
\scalebox{0.6}{\includegraphics{sn78k_fig11.ps}}
\label{theor_obs}
\end{figure*}

SN1978K clearly belongs to this category of behavior.  The Mg\,{\sc
ii} lines originate in the ejecta (detected in the on-source pointing)
which has been decelerated; the narrow lines widths ($\sim$500-900 km
s$^{-1}$) support this picture.  The Ly${\alpha}$ and [Ne~IV] lines
show a larger velocity offset than the Mg\,{\sc ii} lines, but have
similar line velocity widths.  These lines may originate in the
pre-shock medium but the inferred separations from the explosion site
are very large.

The [O~III] lines indicate a high density \citep[Paper I]{R+93,CDD95}
as do the pre-shock circumstellar lines of [N~II] \citep{Chu99}.  The
critical densities of the observed lines also provide clues to the
number density of the emitting gas.  Critical densities range from
[Ne\,{\sc iv}] $\sim$ 4$\times$10$^4$ to [Ne\,{\sc iii}] 3869{\AA} at
$\sim$8$\times$10$^6$ cm$^{-3}$.  Note that no evidence for any
emission near 3727{\AA} is present; the critical density for the
[O\,{\sc ii}] line is $\sim$5$\times$10$^3$ cm$^{-3}$.  The lack of
detectable emission implies that the number density is not less than
$\sim$10$^{4}$ cm$^{-3}$.

The flux ratio of the Mg\,{\sc ii} lines is $\sim$1.8:1.  These lines
originate in transitions from the levels 3p $^2P_{1 \over 2}$ and 3p
$^2P_{3 \over 2}$ to the ground level (3s $^2S_{1 \over 2}$).  The
collisional deexcitation rates $q$ are 3.6$\times$10$^{-7}$ and
7.3$\times$10$^{-7}$ s$^{-1}$ respectively \citep{PP95}.  The
radiative rate A = 2.6$\times$10$^8$ s$^{-1}$ \citep{Mor91}.  Since
collisional deexcitation dominates for ${\tau}n_{\rm e} > A/q$, where
${\tau}$ is the line center optical depth, then ${\tau}n_{\rm e} >
{\sim}10^{14-15}$ cm$^{-3}$.  This density is very high even for a
line center optical depth $>$10$^2$.  The density in the intercloud
wind in the model of \cite{CDD95}Chugai is $\sim$10$^{6-7}$
cm$^{-3}$.  The site of the Mg\,{\sc ii} emission must be in the
shocked region.  The observed emission lines definitely associated
with SN1978K provide density estimates in the range of 10$^{4-14}$
cm$^{-3}$, indicating that we are sampling a large spatial extent of
the region around SN1978K.

The X-ray and radio luminosities are high \citep[Paper I]{R+93}.
\cite{Mon98} argue, based on the classification of
radio light curves, that the fitted value of ${\beta}$ (-0.9
$<~{\beta}~<$ -0.5) for SN1978K better matches normal SN II radio
behavior than the behavior and ${\beta}$ values of SN IIn and SN Ib/c
(${\beta}~<$ -1.1).  Our fit to all of the available radio data now
yields a ${\beta}~{\lesssim}-1.55{\pm}0.35$, placing SN1978K well into
the SN IIn camp.  The high inferred radio flux supports this result.

We could only establish a range for the index $n$ from the X-ray light
curve ($\sim$4 -- 12).  If we use the radio light curve ($n = (2 m - 3
) / ( m - 1)$, with $m = -{\delta} / 3$), we obtain a similar range:
$\sim$3.4 $\lesssim n \lesssim$ 12.6.  This range is similar to values
of $n$ for SN IIn and SN Ib/c and dissimilar to the $n >$ 20 values
typical of the SN IIL \citep{W96}.  We also estimate a mass loss rate
of $\sim$10$^{-4}$ M$_{\odot}$ yr$^{-1}$ using, for example, equation
7 of \cite{Mon98}Montes et al. (1998).  This value is only slightly
less than the estimate in \cite{Mon98}.

The X-ray flux has remained constant \citep[SPC96]{SPC96,this paper}
while the radio flux has declined.  SN1978K's behavior may be typical
of the ``IIn'' subclass of Type II supernovae \citep{S90,F89},
although the time scale for the decline remains to be established.
SN1986J shows a declining X-ray flux \citep{HBCT98} but has an age
approximately equal to that of SN1978K.  SN1988Z is considerably
younger, yet the X-ray and radio fluxes are also declining
\citep{Are99}.  Undoubtedly, the density and extent of the
circumstellar medium dictate the susequent decline time scale.  The
recent supernova SN1997ab provides another example \citep{Sal98} based
upon its optical spectra, although radio and X-ray observations have
not yet been published.

Two models exist in which to interpret the observations: the shell
model (e.g., \cite{CF94,Ter92}) and the cloud model \citep{CDD95}.
Both models explain the overall observational situation; the
differences are in the details.  The shell model postulates a uniform
pre-supernova wind with a power law density distribution
${\rho}_{circ} \sim r^{-s}$, with {\it s} = 1.5-2.  The supernova
ejecta is modeled as a power law distribution with ${\rho} \sim
r^{-n}$, with {\it n} in the range of 8-12.  The cloud model attempts
to reduce the potentially large mass of circumstellar matter the shell
model can imply.  The cloud model naturally allows radiation to leak
out while the shell model may require a flattened geometry plus a
scattering atmosphere to permit radiation to escape.

\begin{figure*}
\caption[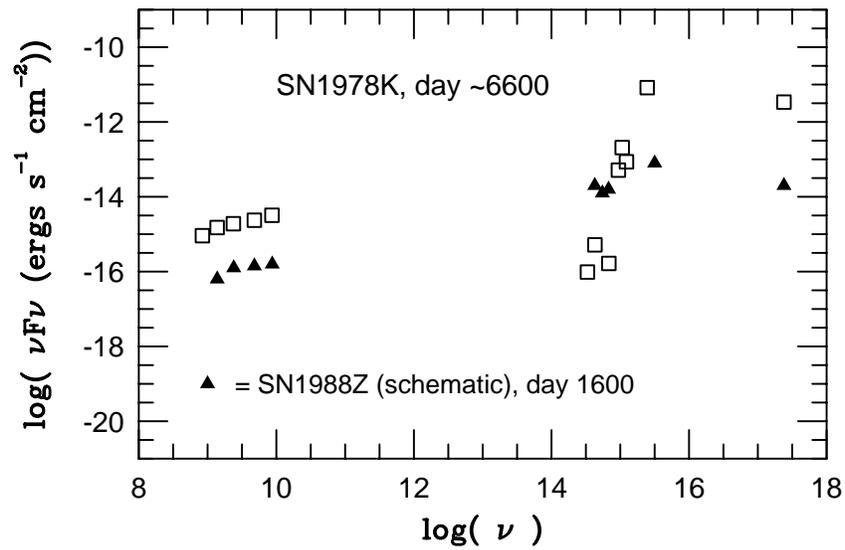]{Spectral energy distribution for SN1978K near
day 6600.  The plotted data comes from an average of the {\it ROSAT}
HRI days 6234 and 7076, from {\it HST} day 6698, from the optical
photometry of day 6595 (B, I) and day 6691 (R), and the radio fluxes
for days 6533 and 6733.} \label{SED}
\scalebox{0.6}{\includegraphics{sn78k_fig12.ps}}
\end{figure*}

The shell model has been used to make specific predictions for
emission lines (\cite{CF94,Ter92}).  Figure~\ref{theor_obs}, and
Table~\ref{comp_obs}, compare the observed emission lines with those
predicted for the shell model by Chevalier \& Fransson (1994)
(``ChevFran'') and Terlevich et al. (1992) (``Ter92'') which uses the
clump model to predict a spectrum for a Narrow Line Region of an AGN.
Figure~\ref{theor_obs} shows the major lines; Table~\ref{comp_obs}
includes lines for which upper limits have been assigned from the
observed spectra.  We have also included the ``B52'' model of
\cite{BDT85}.  This model uses a shock velocity of 86 km s$^{-1}$.
The approximate match of the model to the dereddened line emission
implies that the shock velocity is $\sim$100 km s$^{-1}$, rather than
a high velocity (e.g., 1000 km s$^{-1}$) as is typical of young
supernovae (e.g., \cite{Fes99}), so rapid deceleration must have
occurred in SN1978K.

Neither the clump model nor the shell model provide a good
description of the observed line emission.  Some of the mismatches can
be attributed to the large range of densities that must be present in
the expanding debris of SN1978K: from $\sim$10$^4$ (from [Ne\,{\sc
iv}] 2422{\AA} and [N\,{\sc ii}] 6583{\AA} lines) to $\sim$10$^6$
(from the [Ne\,{\sc iii}] line) to $\sim$10$^{14}$ cm$^{-3}$ (from the
Mg\,{\sc ii} lines).  We also note that the Terlevich et al. model was
generated to describe the spectrum shortly after the explosion and not
to match a spectrum obtained 20 years later.  The large X-ray flux
must also produce emission lines via photoionization.

We can also compare the results from these observations of SN1978K
with recent observations of the Type IIL supernovae SN1979C and
SN1980K \citep{Fes99}.  The X-ray and radio luminosities of the IIL
supernovae are lower by a factor of 10 or more with respect to
SN1978K.  The optical lines are all broad ($\sim$1000-6000 km
s$^{-1}$) in the IIL supernovae.  In addition, no emission lines are
detected that have critical densities below a few $\times$10$^5$
cm$^{-3}$.  On the basis of these results, the IIL supernovae are
clearly undergoing a different evolution.  Fesen et al. conclude that
the shell model is in general agreement with the observations of the
SN IIL remnants.  SN1978K shows generally poor agreement with the
shell model.  We may infer that the shell model is most descriptive of
the SN IIL objects but not the SN IIn's.

A recent paper by \cite{Are99} presents a spectral energy
distribution of the similar SN1988Z.  Figure~\ref{SED} shows the
corresponding distribution for SN1978K near day 6600.  The
distribution looks very similar to that of SN1988Z near day 1600.  The
bulk of the emission is carried by the X-ray and ultraviolet bands as
with SN1988Z.  This suggests that the circumstellar medium surrounding
SN1978K is more dense.  If the evolution time scales directly with the
density, then the medium surrounding SN1978K is at least a factor of
four higher in density.  The other conclusions of the SN1988Z study
apply equally well to SN1978K.

\section{Conclusions}

We have followed the evolution of SN1978K from its discovery in 1992
to the present across the electromagnetic spectrum.  SN1978K's
evolution can be summarized briefly: the X-ray flux is constant, the
optical flux is constant or rising slightly, and the radio flux is
declining.  The constancy of the X-ray flux can not continue,
particularly if the models of dense or compact supernova are at all
correct in predicting rapid evolution.  We intend to continue our
observing program.

\section{Acknowledgements}

We thank the anonymous referee for the comments on the paper;
specifically, we thank the referee for pointing out a reference we had
missed.  We are indebted to E.~Sung, G.~Purcell, and S.~Driver for
acquiring some of the images used for the optical monitoring.  We
thank the {\it HST} FOS team for considerable effort to observe a
challenging object and the Director for approving the 1996
re-observation.  We thank You-Hua Chu and Adeline Caulet for
discussions on the {\it HST} spectra.  EMS thanks I. Salamanca for
comments on an earlier version of the paper.  This research was
supported by Space Telescope grant GO-0536.02-93A and by {\it ROSAT}
grant NAG5-6923 to SAO.

\newpage

\begin{table*}
\begin{center}
\caption{{\it ROSAT} Observations} \label{Obs_table}
\begin{tabular}{lllccccl} 
         &          & Observation &  & Approx. & ExposT &  & Pointing \\
Sequence & Detector &  Date  & MJD$^a$ & Age$^b$ (d)  & (sec) & Flux$^c$ & centered on \\ \hline
150010$^d$ & PSPC & 1990 Jul 23 & 48095 &4445 &  3399 & 1.75$\pm$0.15 &HD~20888 \\
200044$^d$ & PSPC & 1991 Mar 18 & 48333 & 4683 &  3031 & 1.51$\pm$0.19
& HD~20888 \\
600045n00$^{d,e}$ & PSPC & 1991 Apr 24 & 48370 & 4720 & 11118 &
1.99$\pm$0.10 & NGC~1313 \\
400065n00$^{d,f}$ & HRI  & 1992 Apr 18-May 24 & 48730 & 5080 &  5365 &
1.34$\pm$0.22 & NGC~1313 \\
600504n00$^{d,g}$ & PSPC & 1993 Nov 3  & 49299 & 5649 & 15178 &
1.96$\pm$0.09 & NGC~1313 \\
600505ao1$^d$ & HRI  & 1994 Jun 23-Jul 24  & 49538 & 5888 & 22199 &
1.53$\pm$0.10 & NGC~1313 \\
500403n00 & HRI  & 1995 Jan 31-Feb 10  & 49753 & 6103 & 13314 &
2.04$\pm$0.14 & SN~1978K  \\
500404n00 & HRI  & 1995 Feb 2-11     & 49754 & 6104 & 26898 &
1.64$\pm$0.09 & SN~1978K  \\
600505n00 & HRI  & 1995 Apr 12-20    & 49823 & 6173 & 20143 &
1.85$\pm$0.11 & NGC~1313 \\
500403ao1 & HRI  & 1995 May 9-Jul 21 & 49883 & 6233 & 30736 &
1.81$\pm$0.09 & SN~1978K \\
500404ao1 & HRI  & 1995 May 10-Jul 22 & 49884 & 6234 & 18628 &
1.90$\pm$0.12 & SN~1978K \\
500492ao1 & HRI  & 1997 Sep 30-Aug 10 & 50726 & 7076 & 22537 &
1.74$\pm$0.11 & SN~1978K \\ 
500499n00 & HRI  & 1998 Mar 21-Apr 19 & 50908 & 7258 & 23754 &
1.69$\pm$0.10 & SN~1978K \\ \hline
\end{tabular}
\end{center}
Notes:

$^a$MJD at center of observation when spanning multiple days.

$^b$Based on adopted date of maximum of 1978 May 22 = MJD 43650.

$^c$Integrated, unabsorbed flux in 0.2-2.4 keV in units of 10$^{-12}$
ergs s$^{-1}$ cm$^{-2}$ for thermal model with kT=0.4 keV and N$_{\rm
H}$=4.4$\times$10$^{21}$ cm$^{-2}$.

$^d$In light curve presented in \cite[SPC96]{SPC96}.

$^e$Data described in \cite{Col95}.

$^f$Data described in \cite{Stoc95}.

$^g$Data described in \cite{Mill98}.

\end{table*}

\begin{table*}
\begin{center}
\caption{{\it ASCA} Observations} \label{asca_table}
\begin{tabular}{llccc} 
         & Observation &  & Approx$^e$. & On-source \\
Sequence &  Date  & MJD & Age (d) & Time (ksec) \\ \hline
93010000$^a$ & 1993 Jul 12-13 & 49181 & 5531 & 79 \\
60028000     & 1995 Nov 29-30 & 50051 & 6401 & 92 \\ \hline
\end{tabular}
\end{center}
Notes:

$^a$Spectra presented in \cite{Pet94a,Pet94b}.

$^b$Based on adopted date of maximum of 1978 May 22 = MJD 43650.

\end{table*}

\begin{table*}
\begin{center}
\caption{{\it HST} FOS Observations}
\label{point_hist_tab}
\begin{tabular}{llllccc}
       &            &          &             & Exposure  & Total & Aperture \\
 Band$^a$  &  File Name & Location & Offsets$^b$ & Time (sec)& Exp (sec) & Type$^c$ \\ \hline
\multicolumn{7}{c}{1994 September 26 (Age $\sim$5971 d)} \\
 red   & y2em0103t + & NW of SN & -1''.05, +0''.7 & 1980 & 3799 & 1 \\
       & ~~y2em0104t &          &                 & 1819 & & 1 \\
blue   & y2em0107t   & NW of SN & -1''.05, +0''.7 & 3780 & 3780 & 1 \\
 red   & y2em0105t + & SW of SN & -1''.05, -2''.2 & 1720 & 3800 & 1 \\
       & ~~y2em0106t &          &                 & 2080 & & 1 \\
blue   & y2em0108t   & SW of SN & -1''.05, -2''.2 & 3799 & 3799 & 1 \\
\multicolumn{7}{c}{1996 September 22 (Age $\sim$6698 d)} \\
 red   & y30r0205t + & NW of SN & -1''.05, +0''.7 & 1750 & 3800 & 1 \\
       & ~~y30r0206t &          &                 & 2050 & & 1 \\
 red   & y30r0207t + & on SN    & 0.0, 0.0        &  370 & 3949 & 1 \\
       & ~~y30r0208t + &          &                 & 2640 & & 1 \\ 
       & ~~y30r0209t &          &                 &  939 & & 1 \\
 blue  & y30r020bp + & NW of SN & -1''.05, +0''.7 & 1140 & 7820 & 2 \\
       & ~~y30r020ct + &          &                 & 2660 & & 2 \\
       & ~~y30r020dt + &          &                 & 2660 & & 2 \\
       & ~~y30r020ep   &          &                 & 1360 & & 2 \\ \hline
\end{tabular}
\end{center}

Notes:  no entries are listed for acquistion or ``peak up'' exposures.

$^a$The {\it HST} FOS manual refers to the bands as ``blue'' and
``amber''.  Throughout this paper, we use the more familiar ``blue''
and ``red''.

$^b$Offsets are listed as (${\Delta}$RA, ${\Delta}$Dec), in arcsec,
from the position of SN1978K (J2000: $\alpha$ = 03:17:38.7, $\delta$ =
66:3:04.6).

$^c$Aperture types:  1 = single circular, 0''.26 diameter; 2 = pair,
square, 0''.21 on a side;
\end{table*}

\begin{table*}
\begin{center}
\caption{Optical Photometric Monitoring of SN~1978K}
\label{optmags}
\begin{tabular}{cccccccc}
        &  & Approx$^a$. & & & & \\
UT Date  & MJD  & Age   & Telescope  &     $B$        &    $V$         &    $R$         &
    $I$  \\
\hline
1990 Nov 17 & 48212 & 4562 & AAT 3.9m   &  $\ldots$      & $20.23\pm0.28$ & $\ldots$       &
 $20.52\pm0.39$ \\
1992 May 01 & 48743 & 5093 & MSSSO 1.0m &  $\ldots$      & $\ldots$       & $\ldots$       &
 $19.90\pm0.24$ \\
1992 Nov 03 & 48929 & 5279 & MSSSO 1.0m &  $\ldots$      & $19.97\pm0.13$ & $\ldots$       &
 $\ldots$  \\
1994 Jan 09 & 49361 & 5711 & MSSSO 1.0m & $20.89\pm0.28$ & $19.94\pm0.19$ & $18.73\pm0.20$ &
 $19.60\pm0.22$ \\
1995 Feb 24 & 49722 & 6122 & CTIO 1.5m  & $20.80\pm0.26$ & $\ldots$       & $\ldots$       &
 $20.14\pm0.50$ \\
1996 Jan 03 & 50085 & 6435 & $HST^b$      & $20.67\pm0.11$  & $19.81\pm0.05$  & $\ldots$       &
 $\ldots$  \\
1996 Jun 12 & 50246 & 6596 & MSSSO 1.0m & $20.66\pm0.26$ & $\ldots$       & $\ldots$       &
 $19.80\pm0.55$ \\
1996 Sep 15 & 50341 & 6691 & AAT 3.9m   & $\ldots$       & $\ldots$       & $18.49\pm0.29$ &
 $\ldots$ \\
\hline
\end{tabular}
\end{center}

$^a$Age based on adopted date of maximum of 1978 May 22 = MJD 43650.

$^b$pseudo-B and V mags; B from F439W filter, V from F555W filter
\end{table*}

\begin{table*}
\begin{center}
\caption{Optical Line Fluxes and Upper Limits from October 1996 (Age
$\sim$6714 d) AAT Spectrophotometry}
\label{aatlines}
\begin{tabular}{cccrr}
$\lambda$ (\AA) & $\lambda_{0}$ (\AA) &  ID  & Flux &
velocity \\ 
        &      &  &  $({\rm H}\beta=100)$ & (km s$^{-1}$) \\ \hline
$3636.1\pm0.5$ & 3628.6  & [Fe\,{\sc ii}]        &  $27\pm3$ &620$\pm$40 \\
$\ldots$       & 3727    & [O\,{\sc iii}]        & $<$15 & $\ldots$ \\
$3871.1\pm1.1$ & 3868.7  & [Ne\,{\sc iii}]       & $12\pm2$ & 263$\pm$80 \\
$3892.7\pm0.3$ & 3889.0  & He\,{\sc i}, H$\zeta$ & $41\pm1$ & 285$\pm$25 \\
$3973.3\pm0.2$ & 3970.1  & H$\epsilon$, [Ne\,{\sc iii}] & $28\pm3$ & 241$\pm$20 \\
$4033.0\pm4.0$ & 4026.2  & He\,{\sc i}           & $7\pm1$ & 500$\pm$300 \\
$4074.2\pm1.0$ & 4068.6  & [S\,{\sc ii}]         & $21\pm2$ & 410$\pm$70 \\
$4106.2\pm0.2$ & 4101.7  & H$\delta$             & $35\pm5$ & 330$\pm$15 \\
$4248.4\pm2.0$ & 4244.0  & [Fe\,{\sc ii}]        & $13\pm2$ & 310$\pm$140 \\
$4290.5\pm2.0$ & 4287.4  & [Fe\,{\sc ii}]        & $19\pm1$ & 220$\pm$140 \\
$4325.1\pm1.0$ & 4319.6  & [Fe\,{\sc ii}]        & $5\pm2$ & 380$\pm$70 \\
$4345.5\pm0.3$ & 4340.5  & H$\gamma$             & $53\pm2$ & 345$\pm$20 \\
$4364.3\pm0.6$ & $\ldots$  & [O\,{\sc iii}], [Fe\,{\sc ii}] & $16\pm1$ & $\ldots$ \\
$4421.1\pm2.0$ & 4415.0  & [Fe\,{\sc ii}]        & $12\pm1$ &415$\pm$135 \\
$\ldots$       & 4471    & He\,{\sc i}           & $<$15   & $\ldots$ \\
$4578.2\pm0.4$ & 4571.2  & Mg\,{\sc i}]          & 9$\pm$6 & 460$\pm$30 \\
$4691.6\pm0.4$ & 4685.7  & He\,{\sc ii}          & 7$\pm$3 & 380$\pm$30 \\
$4867.2\pm0.2$ & 4861.3  & H$\beta$              & 100$^a$ & 365$\pm$15 \\
$4965.5\pm0.3$ & 4958.9  & [O\,{\sc iii}]        & $14\pm6$ & 400$\pm$20 \\
$5013.2\pm0.3$ & 5006.9  & [O\,{\sc iii}]        & $23\pm3$ & 380$\pm$20 \\ 
$5165.6\pm1.0$ & 5158.8  & [Fe\,{\sc ii}]        & $19\pm5$ & 390$\pm$60 \\
$5760.1\pm1.5$ & 5754.6  & [N\,{\sc ii}]         & $9\pm1$ & 290$\pm$80 \\
$5882.0\pm0.6$ & 5875.7  & He\,{\sc i}           & $20\pm4$ & 320$\pm$30 \\
$6306.6\pm1.5$ & 6300.3  & [O\,{\sc i}]          & $23\pm5$ & 300$\pm$70 \\
$6369.2\pm3.5$ & 6363.8  & [O\,{\sc i}]          & $8\pm2$ & 260$\pm$165 \\
$6569.0\pm0.2$ & 6562.8  & H$\alpha$             & $276\pm10$ & 285$\pm$10 \\
$6589.3\pm0.4$ & 6583.4  & [N\,{\sc ii}]         & $26\pm2$ & 270$\pm$20 \\
\hline
\end{tabular}
\end{center}

$^a$The flux of H${\beta}$ is $\sim$2.0$\times$10$^{-14}$ ergs
s$^{-1}$ cm$^{-2}$ to within 10\%.
\end{table*}

\begin{table*}
\begin{center}
\caption{New Radio Fluxes (mJy) for SN 1978K from ATCA and MOST Monitoring}
\label{atcafluxes}
\begin{tabular}{lccrrrrr}
     &  & Approx.$^a$    & \multicolumn{5}{c}{Band$^b$}                           \\
UT Date  & MJD & Age (d)   & \multicolumn{1}{c}{L} & \multicolumn{1}{c}{S} &
 \multicolumn{1}{c}{C} & \multicolumn{1}{c}{X} &
 \multicolumn{1}{c}{MOST} \\ \hline
1992 Jul 02 & 48805 & 5155 & $139.7\pm1.2$ & $111.4\pm1.6$ & $67.1\pm0.3$ & $43.2\pm0.4$ &
 \multicolumn{1}{c}{$\ldots$} \\
1993 Feb 08 & 49026 & 5376 & $140.7\pm4.5$ & $104.2\pm3.2$ & $61.6\pm0.6$ & $38.7\pm0.4$ &
 \multicolumn{1}{c}{$\ldots$} \\
1993 Oct 24 & 49284 & 5634 & $129.0\pm2.1$ & $97.4\pm1.0$  & $58.4\pm0.5$ & $37.8\pm0.7$ &
 \multicolumn{1}{c}{$\ldots$} \\
1994 Oct 02 & 49627 & 5977 & $118.0\pm4.3$ & $86.6\pm1.8$  & $52.7\pm0.5$ & $33.2\pm0.5$ &
 \multicolumn{1}{c}{$\ldots$} \\
1995 Sep 09 & 49969 & 6319 & $105.5\pm1.9$ & $82.5\pm1.2$  & $48.7\pm0.4$ & $31.7\pm0.3$ &
 \multicolumn{1}{c}{$\ldots$} \\
1995 Nov 08 & 50029 & 6379 &  \multicolumn{1}{c}{$\ldots$} & \multicolumn{1}{c}{$\ldots$} &
 \multicolumn{1}{c}{$\ldots$} &  \multicolumn{1}{c}{$\ldots$}     &
$113\pm8$ \\
1996 Feb 08 & 50121 & 6471 & $120.5\pm6.5$ & $81.5\pm3.7$  & $46.4\pm1.5$ & $30.5\pm0.4$ &
 \multicolumn{1}{c}{$\ldots$} \\
1996 Apr 10 & 50183 & 6533 & $109.2\pm1.3$ & $80.3\pm1.2$  & $49.3\pm0.6$ & $37.1\pm0.4$ &
 \multicolumn{1}{c}{$\ldots$} \\
1996 Oct 30 & 50386 & 6736 & \multicolumn{1}{c}{$\ldots$} &  \multicolumn{1}{c}{$\ldots$} &
 \multicolumn{1}{c}{$\ldots$} &  \multicolumn{1}{c}{$\ldots$} &
$108\pm5$ \\
1997 Aug 31 & 50691 & 7041 & \multicolumn{1}{c}{$\ldots$} &  \multicolumn{1}{c}{$\ldots$} &
 \multicolumn{1}{c}{$\ldots$} &  \multicolumn{1}{c}{$\ldots$} &
$86\pm3$ \\
1998 Aug 25 & 51050 & 7400 &  $88.1\pm4.9$ & $63.5\pm1.6$  & $37.4\pm0.6$ & $24.5\pm0.3$ &
 \multicolumn{1}{c}{$\ldots$} \\
\hline
\end{tabular}
\end{center}

$^a$ Age based on adopted date of maximum of 1978 May 22 = MJD 43650.

$^b$ Central frequencies are 1380, 2370, 4790, and 8640~MHz for
L, S, C, and X-bands respectively, except for 1998 Aug 25, when
the central S-band frequency was 2496~MHz. MOST central frequency
is 843~MHz.

\end{table*}

\begin{table*}
\begin{center}
\caption{Model Fits to {\it ASCA} Spectra} \label{asca_fits}
\begin{tabular}{lccccc}
      & Index or  &   N$_{\rm H}$\tablenotemark{a}  &
\multicolumn{2}{c}{Unabsorbed Flux\tablenotemark{b}} &  \\
Label & Temp (keV)\tablenotemark{a} & 
(10$^{22}$ cm$^{-2}$) & 0.5-2.0 keV & 2.0-10.0 keV &
${\chi}^2$/dof \\ \hline
\multicolumn{6}{c}{1993 July 12-13} \\ 
Power  & 3.06$^{+0.74}_{-0.33}$ & 0.28$^{+0.15}_{-0.10}$ & 1.4(-12) & 3.3(-13) & 45.1/41 \\
Brems  & 1.27$^{+0.93}_{-0.41}$  & 0.14$^{+0.10}_{-0.07}$ & 8.0(-13) & 2.4(-13) & 44.0/41 \\
Mekal-frozen\tablenotemark{c} & 0.68$^{+0.11}_{-0.07}$ &
0.76$^{+0.10}_{-0.12}$ & $\ldots$ & $\ldots$ & 42.3/41 \\
Mekal-thaw\tablenotemark{d} & 0.83$^{+0.90}_{-0.05}$ &
0.25$^{+0.10}_{-0.15}$ & 1.1(-12) & 1.4(-13) & 39.2/39 \\
\multicolumn{6}{c}{1995 November 29-30} \\
Power  & 2.81$^{+0.35}_{-0.30}$ & 0.26$^{+0.07}_{-0.06}$ & 1.4(-12) & 4.8(-13) & 75.1/68 \\
Brems  & 1.27$^{+1.50}_{-0.20}$ & 0.16$^{+0.03}_{-0.11}$ & 9.2(-13) & 2.8(-13) & 71.3/68 \\
Mekal-frozen\tablenotemark{c} & 0.63$^{+0.08}_{-0.07}$ & 0.79$^{+0.06}_{-0.07}$ & $\ldots$ & $\ldots$ &
74.2/68 \\
Mekal-thaw\tablenotemark{d} & 0.71$^{+1.10}_{-0.03}$  & 0.37$^{+0.04}_{-0.20}$ & 1.6(-12) & 1.3(-13) &
62.9/66 \\ \hline
\end{tabular}
\end{center}

\tablenotetext{a}{error bars are 1 $\sigma$}

\tablenotetext{b}{units = ergs s$^{-1}$ cm$^{-2}$}

\tablenotetext{c}{Abundance frozen at 1.0.}

\tablenotetext{d}{Abundance thawed; epoch 1 value = 0.04; epoch 2
value = 0.05; both are identical within the errors}

\end{table*}

\begin{table*}
\begin{center}
\caption{Fits to Emission Lines: {\it HST} Observation} \label{FOS_lines}
\begin{tabular}{llccccccccr}
  &  & \multicolumn{2}{c}{Observed} & \multicolumn{2}{c}{Dereddened Flux\tablenotemark{b}} & \multicolumn{2}{c}{Luminosities\tablenotemark{b}} & Rest     & Radial  & \\
Location\tablenotemark{a} & Line & ${\lambda}$ & Flux     & Low 
&High   & Low & High & ${\lambda}$ & Velocity & FWHM \\ 
 & & ({\AA}) & \multicolumn{3}{c}{(10$^{-14}$ erg s$^{-1}$ cm$^{-2}$)} &
\multicolumn{2}{c}{(10$^{37}$ erg s$^{-1}$)} & ({\AA}) & (km s$^{-1}$) & ({\AA}) \\ \hline
\multicolumn{11}{c}{1994 September 26 (Age $\sim$5971 d)} \\
NW & Ly-${\alpha}$ & 1219.0 & 2.2 & 2.4 & 52. & 7.4 & 160. & 1215.7 & 830 & 6.40 \\
NW & [Ne~IV] & 2431.6 & 0.05 & 0.05 & 0.48 & 0.15 & 1.3 & 2422.3 & 1150 & 9.45\\
SW & Ly-${\alpha}$ & 1219.3 & 3.6 & 4.0 & 86. & 12. & 270. & 1215.7 & 890 & 6.68 \\
SW & [Ne~IV] & 2433.2 & 0.03 & 0.04 & 0.16 & 0.10 & 0.32 & 2422.3 & 1350 & 5.86\\
\multicolumn{11}{c}{1996 September 22 (Age $\sim$6698 d)} \\
NW & Ly-${\alpha}$ & 1220.4 & 2.4 & 2.7 & 57. & 8.4 & 180. &1215.7 & 1170 & 4.09 \\
NW & [Ne~IV] & 2435.1 & 0.03 & 0.04 & 0.29 & 0.10 & 0.96 & 2422.3 & 1600 &11.3 \\
SN & Mg II  & 2801.4 & 0.03 & 0.03 & 0.16 & 0.09 & 0.47 & 2795.5 & 630 & 3.4\tablenotemark{c} \\
SN & Mg II  & 2808.7 & 0.02 & 0.02 & 0.11 & 0.05 & 0.26 & 2802.7 & 640& 3.4\tablenotemark{c} \\ 
SN & He I  & 3190.2 & 0.01 & 0.01 & 0.05 & 0.03 & 0.29 & 3187.7\tablenotemark{d} & 240
& 6.2 \\ \hline
\end{tabular}
\end{center}

Notes:

\tablenotetext{a}{Location:  SN = centered on SN; NW, SW = offset from SN}

\tablenotetext{b}{Two values used for E$_{\rm {B-V}}$ because
\cite{R+93} found E$_{\rm {B-V}}$ $\sim$0.31, while the 1996 AAT
optical spectrum imples E$_{\rm {B-V}}$ is very low (perhaps as low as
the \cite{BH84} value of E$_{\rm {B-V}}$ = 0.01).}

\tablenotetext{c}{FWHM was fit, but constrained to be identical for
both lines.}

\tablenotetext{d}{Assuming the He I identification is correct.}
\end{table*}

\begin{table*}
\begin{center}
\caption{Upper Limits to Emission Lines:  {\it HST} Observation} \label{FOS_UL}
\begin{tabular}{lcccc}
       &    Rest         & Flux$^b$    & \multicolumn{2}{c}{Dereddened Flux$^b$} \\
 Line  &  Wavelength$^a$ & Upper Limit & Low    & High \\ \hline
  N V  & 1240  & 5(-16) & 5.6(-16) & 1.2(-14) \\
 O I   & 1304  & 6(-16) & 6.5(-16) & 9.5(-15) \\
 SiIV/O IV  & 1400  & 5(-16) & 5.5(-16) & 6.5(-15) \\
 C IV  & 1550  & 4(-17) & 4.3(-17) & 4.1(-16) \\
He II  & 1640  & 3(-16) & 3.2(-16) & 2.9(-15) \\
C III  & 1909  & 3(-16) & 3.2(-16) & 3.2(-15) \\
C II$]$& 2324  & 2(-16) & 2.2(-16) & 1.9(-15) \\
Si II  & 2335  & 7(-17) & 7.6(-17) & 6.6(-16) \\
$[$O II$]$ & 2470 & 3(-17) & 3.2(-17) & 2.9(-16) \\
  He II & 2733 & 5(-17) & 5.3(-17) & 3.1(-16) \\
 C I    & 2967 & 3(-17) & 3.2(-17) & 2.0(-16) \\ \hline
\end{tabular}
\end{center}

$^a$ units = {\AA}

$^b$ units = ergs s$^{-1}$ cm$^{-2}$; notation defined as N(-XX) = N$\times$10$^{-XX}$
\end{table*}

\begin{table*}
\begin{center}
\caption{Comparison of Balmer Decrement in Optical Spectra} \label{Balm_decr}
\begin{tabular}{crrrrrr}
     & \multicolumn{4}{c}{Observation Date} \\
 Ratio  & Jan 1990\tablenotemark{a}   & Mar 1992\tablenotemark{a}  & Oct 1996   & Case A\tablenotemark{b}  & Case B\tablenotemark{c} \\ \hline
 H${\alpha}$/H${\beta}$ & 476$\pm$54 & $\ldots$ &276$\pm$10 & 3.10 & 2.81  \\
 H${\beta}$/H${\beta}$ & 100 & 100 & 100  & $\ldots$ & $\ldots$ \\
 H${\gamma}$/H${\beta}$ & 38$\pm$6 & 46$\pm$2 & 53$\pm$2 & 0.46 & 0.47 \\
 H${\delta}$/H${\beta}$ & 19$\pm$2 & 16$\pm$4 & 35$\pm$5 & 0.25 & 0.26 \\
 H${\epsilon}$/H${\beta}$ & 16$\pm$1 & 12$\pm$1 & 28$\pm$3 & 0.15 &
 0.16 \\ \hline
\end{tabular}
\tablenotetext{a}{\cite{R+93}}

\tablenotetext{b}{Assumes T = 5000 K.  For T = 10,000 K, values are
nearly identical to Case B.}

\tablenotetext{c}{Assumes T = 10,000 K and N$_{\rm e}$ = 10$^6$ cm$^{-3}$.}
\end{center}
\end{table*}

\newpage
\begin{deluxetable}{lccccccc}
\tablecaption{Comparison of Radio Light Curve Model Parameters} \label{mwpar}
\tablehead{
\colhead{Parameter\tablenotemark{a}} & 
\colhead{SN 1978K\tablenotemark{b}}  & 
\colhead{SN 1978K\tablenotemark{c}}  & 
\colhead{SN 1986J\tablenotemark{d}}  & 
\colhead{SN 1988Z\tablenotemark{e}}  & 
\colhead{SN 1993J\tablenotemark{f}}  & 
\colhead{SN 1979C\tablenotemark{g}}  & 
\colhead{SN 1980K\tablenotemark{h}}
}
\startdata
$K_{1}$ (mJy)  & $(0.2-76)\times10^{7}$   & $(2.1-38)\times10^{4}$  &$(3.8-9.2)\times10^{5}$  
  & $(6.6-11.5)\times10^{4}$  & 1.81-6.25$\times$10$^3$ & (1.1-1.8)$\times$10$^3$ & 81-168 \\

$\alpha$ & $-(0.77\pm0.01)$  & $-(0.81-0.73)$  & $-(0.59-0.71)$ 
  & $-(0.69-0.78)$   & -(1.60-0.67) & -(0.66-0.79) & -(0.56-0.67) \\

$\beta$  & $-(1.55\pm0.35)$  & $-(1.0-0.69)$  & $-(1.16-1.22)$  
  & $-(1.43-1.47)$  &  $-(0.89-0.53)$  & -(0.76-0.81)   & $-(0.78-0.68)$ \\

$K_{2}$  & $(0.3-90)\times10^{4}$   & $(1.8-16)\times10^{7}$  &$(0-63)\times10^{5}$  
& $(0-20.6)\times10^{5}$ & (0.72-7.65)$\times$10$^3$ &(2.4-6.8)$\times$10$^7$ & (0.60-15)$\times$10$^5$ \\

$\delta$ & $-(2.22\pm0.36)$  & $-2.92$ (fixed)  & $-(2.19-2.69)$  
  & $-(2.22-2.35)$ & -(2.15-1.61) & -(2.85-3.03)  & -(2.12-3.11)   \\

$K_{3}$  & $(0.03-30)\times10^{11}$ & $<4.2\times10^{12}$ & $(2-12)\times10^{12}$ 
&$(4.0-11.1)\times10^{11}$  &  $(1.46-7.65)\times10^{4}$  & $\ldots$  & $\ldots$  \\

$\delta^{\prime}$  & $-(3.7\pm0.6)$  &  $-4.87$ (fixed)  &  $-(3.65-4.45)$  
  &  $-(3.70-3.92)$ &$-(2.18-1.71)$   &  $-4.93$  &  $\ldots$  \\

$K_{4}$  & $(9\pm1)\times10^{-3}$  & $(7.6-12)\times10^{-3}$  & $\ldots$
  & $\ldots$   & $\ldots$ & $\ldots$ & $\ldots$ \\

$L_{5~{\rm GHz}}^{\rm peak}$  & $(0.9-2.0)\times10^{28}$ &$6.1\times10^{27}$   & $1.7\times10^{28}$  
  & $2.1\times10^{28}$ & 1.1$\times$10$^{30}$ & 8.8$\times$10$^{29}$ & 7.6$\times$10$^{28}$  \\
\enddata

\tablenotetext{a}{L$_{5~{\rm GHz}}^{\rm peak}$ = Model peak 5~GHz luminosities:
Units are ergs s$^{-1}$ Hz$^{-1}$. 
Luminosity values for SN~1986J and SN~1988Z are
taken from the revised values in Montes, Weiler, \& Panagia (1997).}

\tablenotetext{b}{This paper.}

\tablenotetext{c}{\cite[MWP]{MWP}}

\tablenotetext{d}{\cite{W90}}

\tablenotetext{e}{\cite{VD93}}

\tablenotetext{f}{\cite{VD94}}

\tablenotetext{g}{\cite{W91}}

\tablenotetext{h}{\cite{Mon98}}
\end{deluxetable}

\newpage

\begin{table*}
\begin{center}
\caption{Comparison of Predictions with Observed Line Strengths\tablenotemark{a}} 
\label{comp_obs}
\begin{tabular}{lrrrrr}
 Line       & NLR\tablenotemark{b} & CF94\tablenotemark{c}  &
BDT85\tablenotemark{d} & 1992\tablenotemark{e} & 1996\tablenotemark{f} \\ \hline
Ly$\alpha$    & 24.9  & 74.6 & $\ldots$ & $\ldots$ & 9.90 \\
2325 C\,{\sc ii}]    &  0.25 &  3.08 & $\ldots$ & $\ldots$ &  $<$0.03 \\
2422 [Ne\,{\sc iv}]  &  0.001 & 0.52 & $\ldots$ & $\ldots$ &  0.05 \\
2800 Mg\,{\sc ii}    &  1.14 & 17.05 & $\ldots$ & $\ldots$ &  0.05 \\
3869 [Ne\,{\sc iii}] &  1.40 &  1.62 & 0.06 & $\ldots$ &  0.15 \\
4069 [S\,{\sc ii}]   &  0.04 &  1.04 & 0.22 & $\ldots$ &  0.26 \\
4363 [O\,{\sc iii}]  &  0.26 &  0.0  & 0.07 & $\ldots$ &  0.18 \\
4861 H$\beta$        & 1.0   &  1.0  & 1.0 & 0.41 & 1.0 \\
4959+5007 [O\,{\sc iii}]  & 22.3  & 13.75 & 1.37 & 0.15 & 0.35 \\
5876 He\,{\sc i}     &  0.10 &  0.0  & 0.09 & 0.06 & 0.16 \\
6300/6363 [O\,{\sc i}] & 0.60 & 2.13 & 0.57 & 0.16 &  0.23 \\
H$\alpha$     &  2.75 &  2.98 &  3.0 & 2.90 & 1.98 \\
6584 [N\,{\sc ii}]   &  0.93 &  2.22 & 1.92 & 0.21 & 0.19 \\ \hline
\end{tabular}

\tablenotetext{a}{All values are scaled to dereddened H$\beta$ = 1.0 (=
5.7$\times$10$^{-14}$ ergs s$^{-1}$ cm$^{-2}$).}

\tablenotetext{b}{NLR = Narrow Line Region prediction of Terlevich et
al. (1992).}

\tablenotetext{c}{CF94 = \cite{CF94}}

\tablenotetext{d}{BDT85 = \cite{BDT85}, model B52}

\tablenotetext{e}{1992 observation of \cite{CDD95}}

\tablenotetext{f}{{\it HST} and AAT data,using `high' reddening
correction to obtain the most conservative comparison}
\end{center}
\end{table*}

\clearpage

%
\end{document}